\documentclass[usenatbib]{mn2e}
\usepackage{url} 
\usepackage{amsmath}
\usepackage{epsfig}
\usepackage{graphicx}
\usepackage{ulem}
\usepackage{color}

\newcommand\lsim{\mathrel{\rlap{\lower4pt\hbox{\hskip1pt$\sim$}}
        \raise1pt\hbox{$<$}}}
\newcommand\gsim{\mathrel{\rlap{\lower4pt\hbox{\hskip1pt$\sim$}}
        \raise1pt\hbox{$>$}}}

\newcommand{\K}{\mathrm{K}}

\newcommand{\g}{\mathrm{g}}

\newcommand{\dd}{\partial}

\newcommand{\cm}{\;\mathrm{cm}}

\newcommand{\s}{\;\mathrm{s}}

\newcommand{\km}{\;\mathrm{km}}
\newcommand{\yr}{\;\mathrm{yr}}

\newcommand{\Hz}{\;\mathrm{Hz}}

\newcommand{\pc}{\;\mathrm{pc}}
\newcommand{\Mpc}{\;\mathrm{Mpc}}

\newcommand{\Msol}{M_{\odot}}

\newcommand{\Mch}{\mathcal{M}}
\newcommand{\apj}{ApJ}
\newcommand{\apjl}{ApJ}
\newcommand{\apjs}{ApJS}
\newcommand{\aap}{A$\&$A}
\newcommand{\araa}{ARAA}
\newcommand{\actaa}{ACTAA}
\newcommand{\mnras}{MNRAS}
\newcommand{\pasj}{PASJ}
\newcommand{\prd}{PRD}
\newcommand{\aj}{AJ}
\newcommand{\nat}{Nature}

\defcitealias{SesVec10}{SV10}
\defcitealias{CorCor10}{CC10}
    \def\dd{\partial}
    \def\tilde{\widetilde}

    \def\beq{\begin{equation} }
    \def\eeq{\end{equation} }
    \def\spose#1{\hbox to 0pt{#1\hss}}
    \def\ltsim{\mathrel{\spose{\lower.5ex\hbox{$\mathchar"218$}}
     \raise.4ex\hbox{$\mathchar"13C$}}}

\def\tilde{\widetilde}
\def\spose#1{\hbox to 0pt{#1\hss}}
\def\lta{\mathrel{\spose{\lower 3pt\hbox{$\mathchar"218$}}
        \raise 2.0pt\hbox{$\mathchar"13C$}}}
\def\gta{\mathrel{\spose{\lower 3pt\hbox{$\mathchar"218$}}
        \raise 2.0pt\hbox{$\mathchar"13E$}}}

\title[Electromagnetic counterparts of PTA sources]
{Electromagnetic counterparts of supermassive black hole binaries resolved by pulsar timing arrays}

\author[Tanaka, Menou \& Haiman]{
Takamitsu Tanaka$^{1,2}$\thanks{E-mail:taka@mpa-garching.mpg.de},
Kristen Menou$^{1}$, and Zolt\'an Haiman$^{1}$
\\
$^{1}$Department of Astronomy, Columbia University, 550 West 120th Street, New York, NY 10027, USA
\\
$^{2}$Max Planck Institute for Astrophysics, Karl-Schwarzschild-Str.~1, D-85741 Garching, Germany
}

\begin{document}
\maketitle

\label{firstpage}

\begin{abstract}
  Pulsar timing arrays (PTAs) are expected to detect gravitational
  waves (GWs) from individual low--redshift ($z\ltsim 1.5$) compact
  supermassive ($M\gta 10^{9}\Msol$) black hole (SMBH) binaries with
  orbital periods of $\sim 0.1 - 10 \yr$.  Identifying the
  electromagnetic (EM) counterparts of these sources would provide
  confirmation of putative direct detections of GWs, present a rare
  opportunity to study the environments of compact SMBH binaries, and
  could enable the use of these sources as standard sirens for
  cosmology.  Here we consider the feasibility of such an EM
  identification.  We show that because the host galaxies of resolved
  PTA sources are expected to be exceptionally massive and rare, it
  should be possible to find unique hosts of resolved sources out to
  $z\approx 0.2$.  At higher redshifts, the PTA error boxes are
  larger, and may contain as many as $\sim 100$ massive-galaxy
  interlopers.  The number of candidates, however, remains tractable
  for follow-up searches in upcoming wide-field EM surveys. We develop
  a toy model to characterize the dynamics and the thermal emission
  from a geometrically thin gaseous disc accreting onto
  a PTA-source SMBH binary.
  Our model predicts that at optical and infrared frequencies,
  the source should appear similar to a typical luminous active galactic nucleus (AGN).
  However, owing to the evacuation of the accretion flow
  by the binary's tidal torques, the source might have an
  unusually low soft X-ray luminosity
  and weak UV and broad optical emission lines,
  as compared to an AGN powered by a single SMBH
  with the same total mass.
  For sources at $z\sim 1$, the decrement in the rest-frame UV
  should be observable as an extremely red optical color.
  These properties would make the PTA sources
  stand out among optically luminous AGN,
  and could allow their unique identification.
  Our results also suggest that accreting
  compact SMBH binaries may be included among the observed
  population of optically bright, X-ray-dim AGN.
\end{abstract}

\section{Introduction}
\label{sec:intro}

Over the last several years, the possibility of observing both the
gravitational-wave (GW) and electromagnetic (EM) emission signatures
of coalescing supermassive black hole (SMBH) binaries has received
intense attention (\citealt{HH05, Kocsis+06, Kocsis+07, Kocsis+08,
  Dotti+06};
for specific proposed mechanisms for EM signatures,
see \citealt{AN02} and \citealt{MP05}, as well as recent reviews by
\citealt{Haiman+09} and \citealt{Schnittman11}).
The bursts of GWs emitted by such systems can now be predicted by
numerical general relativity \citep{Pretorius05, Baker+06, Campan+06},
and are expected to be observed by current and future detectors.
The temporal evolution of the gravitational
waveform can be used to extract the luminosity distance, help
constrain the location of the source on the sky, and determine the
masses and spins of the SMBHs.  If an EM signature of the coalescence
can also be identified, this would allow for a determination of the
source redshift, turning merging black holes into ``standard sirens''
for probing cosmic expansion.\footnote{ The importance of such GW+EM
  observations for cosmography was first discussed by \cite{Schutz86}
  in the context of merging neutron star binaries.}  Such
multi-messenger observations would also enable astronomical
investigations of SMBHs whose masses, spins and orbital parameters are
already known, presenting ideal laboratories for investigating
accretion physics in active galactic nuclei (AGN).  Furthermore, if
major mergers of galaxies trigger luminous AGN activity (e.g.,
\citealt{Sanders+88, Hernquist89, Carlberg90, BarnesHernquist91,
  HernquistMihos95, MihosHernquist96, KauffmannHaehnelt00,
  Hopkins+07a, Hopkins+08}), then the characteristic EM emission
promptly following the SMBH coalescence may herald the birth of a
quasar \citep{THM10}.

To date, theoretical studies of EM signatures of GW-emitting SMBH
binaries have largely centred on systems predicted to be detectable
by future space-based laser interferometers such as
{\it Laser Interferometer Space Antennae} ({\it LISA}),
with total mass $\sim 10^{5-7}(1+z)^{-1}\Msol$ and redshifts as high as $z\gta 10$.
For the purposes of multi-messenger astronomy, 
a paramount feature of interferometer-detectors is the
precision with they could determine the angular sky position of
SMBH sources \citep{Kocsis+06}: to $\ltsim 1 \deg$ \citep{Vecchio04,
  LH08}, or perhaps even to $\ltsim 1^{\prime}$
when spin-induced precession \citep{LH06} or higher-order harmonics
\citep{McWill+10} are included in the analysis of the waveform.

In this paper, we evaluate the prospects of performing multi-messenger
using pulsar timing arrays (PTAs), which aim to
detect low-frequency (nHz) GWs through precision timing
of Galactic millisecond pulsars.
A major goal for PTAs is to detect the stochastic GW background
due to the population of compact SMBH binaries in our cosmic neighborhood ($z\ltsim 2$).
\cite{Jenet+05} showed that this requires the timing of some $\ga 20$
pulsars to $\sim 100 ~{\rm ns}$ precision over 5 years,
a goal that could be achieved by currently operational arrays
\citep[][and refs. therein]{Manchester08, Verbiest+09}.
Recent theoretical population-synthesis studies 
\citep{SVV09, SesVec10, KocSes11}
have predicted that PTA observations may also be able to
{\it individually resolve} the most massive and/or
most nearby SMBH binaries that stick out above
the stochastic background.\footnote{In principle, PTAs can also detect GW-memory bursts from
individual major SMBH mergers \citep{Seto09, Pshirkov+10,HL10}.
However, the number of detections is expected to be quite low,
perhaps much lower than unity over a full observation campaign \citep[see last paragraph in][]{Seto09}.
We therefore limit the present paper to the
population and EM signatures of compact SMBH binaries.}
The primary factor that determines the number of detectable
individual sources is expected to be the array sensitivity, with
theoretical uncertainties such as the binary mass function
and orbital evolution affecting predictions of the observable
population by factors of unity \citep{Sesana+11}.

The systems individually detected by PTAs will
differ from {\it LISA} targets in several ways.
First, according to the population synthesis studies,
the binaries will have chirp masses
\beq
\Mch\equiv M_{1}^{3/5}M_{2}^{3/5}M^{-1/5}=\eta^{3/5}M,
\label{eq:Mch}
\eeq
typically around $\Mch\sim10^{8.5}\Msol$.
Above, $M_{2}\le M_{1}$ are
the masses of each member of the binary,
and $\eta\equiv M_{1}M_{2}/M^{2}\le 1/4$ is the symmetric mass ratio.
In other words, the systems of interest here
are much more massive than those relevant for {\it LISA}.
Second, PTA sources lie at much closer cosmological distances.
While the redshift probability distribution is poorly known,
it is expected to decline steeply outside the range
$0.1\ltsim z\ltsim 1.5$.
The low cutoff is due to the smaller volume contained
within $z$, whereas the high cutoff is due to the attenuation
of the GW signal and the decline of the intrinsic SMBH merger rate.
Third, the PTA targets are not as compact, having periods of 
$P\sim 1\yr$. The larger separations correspond to slower
orbital decay; the vast majority of PTA sources will not
coalesce within a human lifetime.
That the orbit --- and thus the waveform --- is expected to
evolve much less appreciably with time poses a particularly
difficult challenge for determining the masses and luminosity
distances of PTA sources.
Particularly important for the prospects of performing synergistic
EM observations on these sources is their
relatively poor sky-localization.
The solid-angle uncertainty in the source position may be
anywhere from $\Delta \Omega \ltsim 3\deg^{2}$
(in the best-case scenario in which the contributions
to the signal from individual pulsars are known;
\citealt{CorCor10}, hereafter CC10)
to as large as $\Delta \Omega \sim 40\deg^{2}$ (if individual pulsar
contributions cannot be extracted from the data;
\citealt{SesVec10}, hereafter SV10).

Several tell-tale EM features of such compact SMBH binaries
have been proposed in the literature,
including periodic emission modulated at the orbital frequency of the
PTA source (e.g., \citealt{Haiman+09} and references therein) and
double-peaked broad emission lines \citep[e.g.,][]{Gaskell96, Zhou+04,
  Bogdan+08, BlechaLoeb08, BorLau09, Bogdan+09b, Dotti+09,TanGri09,
  ShenLoeb10}.
In this work, we will explore in detail the suggestion by
\cite{MP05} that tidal torques of a near-merger SMBH binary will have
evacuated the central region of its accretion disc,
resulting in a markedly softer thermal spectrum.

Below, we investigate whether individually resolved
PTA sources may be viable targets for EM identification.
In particular, we address the following two questions:

\begin{enumerate}

\item What is the average number $N_g$ of candidate host galaxies ---
  that is, interloping galaxies that could plausibly harbour the GW
  source --- in a typical error box of a PTA detection?  Of particular
  interest is whether there are plausible scenarios for detecting
  individually resolved sources with $N_g<1$, i.e., cases where the
  source may be uniquely identified with an EM search of the
  three--dimensional PTA error box.

\item In cases where $N_g>1$, what can be done to distinguish the true
  host galaxy of the source from the other interlopers?  Motivated by
  the hypothesis that galaxy mergers can fuel AGN activity, we will
  consider the differences in thermal emission properties predicted by
  disc models of AGN powered by a compact SMBH binary as opposed to one
  powered by a solitary SMBH of the same total mass.

\end{enumerate}

This paper is organized as follows.  In \S \ref{sec:hosts}, we provide
a brief overview of the expected population of PTA-resolved SMBH
binaries as well as the anticipated detection error box of such
objects.  We consider the error box prescriptions of
\citetalias{SesVec10} and \citetalias{CorCor10}, characterize the
types of astronomical objects that are plausible hosts of a
PTA-resolved binary, and estimate the number of such objects.  In \S
3, we describe a toy model to calculate the dynamical state and
thermal emission features of gas accreting onto a resolved SMBH
binary.  We discuss several features predicted by the model which, if
observed, could help the EM identification of individually resolved
PTA sources.
We summarize our findings and offer our conclusions in \S 4.

Throughout this paper, $c$ denotes the speed of light;
$G$ is the gravitational constant;
$k_{\rm B}$ is the Boltzmann constant;
$\sigma_{\rm SB}$ is the Stefan-Boltzmann constant;
and $m_{\rm p}$ is the mass of the proton.

\section{Plausible Hosts of PTA-resolved Binaries}
\label{sec:hosts}

As stated in \S \ref{sec:intro}, theoretical models predict that SMBH
binaries individually resolved by PTAs are most likely to have masses
of $M\gta 10^{8}\Msol$, observed periods of $P_{\rm obs}\sim 1 \yr$,
and lie in a redshift range $0.1\ltsim z\ltsim 1.5$.  Given the PTA
detection of such a GW-source binary, we wish to evaluate the number
$N_g$ of candidate galaxies that could plausibly host it.  To this
end, we will first review the volume of the error box in which we must
look for the source, based on previous work on the source localization
capability of PTAs.  Then, we will evaluate the number of interloping
host galaxies in the error box by estimating the number of (i)
sufficiently massive dark matter halos, (ii) sufficiently luminous
luminous galaxies, and (iii) AGN.  We adopt a standard $\Lambda$CDM
cosmology with dimensionless Hubble parameter $h=0.70$,
matter density $\Omega_{\rm m}=0.27$,
dark energy density $\Omega_{\Lambda}=1-\Omega_{\rm m}$,
baryon density $\Omega_{\rm b}=0.046$,
and power spectrum amplitude $\sigma_{8}=0.81$.
\citep[WMAP 7-year results,][]{Jarosik+11}.

\subsection{The PTA Error Box}

We consider two different estimates of the size of the error box of
PTA-resolved sources.  The first is based on the calculations by
\citetalias{SesVec10}, who assumed that the contributions to the signal
from GW perturbations at the individual pulsars (the so-called
``pulsar term'') cannot be extracted from the PTA data. 
Those authors found that for a hypothetical
array containing 100 pulsars and covering the whole sky,
a typical resolved source could be 
be localized within a solid-angle error of $\Delta
\Omega \sim 40~({\rm SNR}/10)^{-2}\deg^{2}$, where SNR is the
signal-to-noise ratio.
The fractional error of the signal amplitude $A\propto
\mathcal{M}^{5/3}D_{L}^{-1}$ will be of order $\sim 30~({\rm SNR}/10)^{-1}\%$.
These values are statistical means from their Monte-Carlo simulations;
the localization and amplitude measurement for an individual
source may be better or worse, depending on the specific
parameters of the binary and the orientation of the pulsars with
respect to the source.
Given the wide spread in chirp mass distribution predicted by population
synthesis models of resolved sources, in the absence of an independent
measurement of $\mathcal{M}$ the only constraint on $D_{L}$ comes from
the maximum distance at which PTAs are expected to detect individually
resolved sources.  The population synthesis models \citep{SVV09,
  KocSes11} predict that the majority of resolved sources will lie
below a maximum redshift $z_{\rm max}\sim 1.5$, or a luminosity
distance below $D_{L, \max}\ltsim 10^{4}\Mpc$.  This ``worst-case''
error box has a comoving volume of
\beq
\Delta V^{\rm (SV)}
\sim
3\times 10^{8} \left(\frac{\Delta \Omega}{40 \deg^{2}}\right)\Mpc^{3}.
\label{eq:SV}
\eeq

More optimistic numbers are obtained by \citetalias{CorCor10}, who
suggested that utilizing information on the distances to individual
pulsars in the array can greatly enhance the measurement capabilities
of PTAs.  They concluded that if the individual pulsar term
can be extracted from the signal, then this would double
the signal power and enable direct measurement of the chirp mass.
They estimate that for a system with SNR$=20$ (corresponding to a
detection of SNR$=10$ without pulsar distance information), a resolved
source can be localized with distance and angular errors of $\Delta
D_{L}/D_{L}<20\%$ and $\Delta \Omega <3 \deg^{2}$, respectively.
Noting that the comoving distance $D(z)$ in the relevant redshift
range can be analytically approximated\footnote{ This fitting formula
  has an error of less than $1\%$ in $D$ at $z\le1.4$ and roughly
  $5\%$ at $z=1.9$.  It is provided for the reader's convenience; all
  distance and volume calculations in this paper are performed using
  exact expressions.  } as $D\approx c H_{0}^{-1}z~(1-0.2z)$, we may
estimate the error box in the \citetalias{CorCor10} scenario as
\beq 
\Delta V^{\rm (CC)} \approx 1.2\times 10^{6}~z^{3}(1-0.2z)^{3}
\left(\frac{\Delta \Omega}{3 \deg^{2}} \right)
\left(\frac{\Delta D_{L}/D_{L}}{20\%} \right) \Mpc^{3}.
\label{eq:CC}
\eeq

\subsection{Interloper Counts}

Where does an individually resolved PTA source live?  The mass $M$ of
a nuclear SMBH is known to correlate with the velocity dispersion
$\sigma$ of the host galaxy \citep[the ``$M-\sigma$
relation'';][]{FerrareseMerritt00, Gebhardt+00, Tremaine+02}, as well
as with the stellar luminosity of the host \citep[the ``$M-L$
relation'';][]{KR95, Magorr+98, HaringRix04, Lauer+07}; with more
massive halos and luminous galaxies hosting more massive SMBHs.  That
resolved PTA sources are expected to be exceptionally massive ($M\gta
10^{8}\Msol$) implies that the host should be a giant elliptical
galaxy or be among the most massive spiral galaxies
(with velocity dispersion $\sigma\ga 200 \km\s^{-1}$ of the spheroid component; e.g.\citealt{Gultekin+09}).

Assuming that SMBH binaries are able to overcome the ``final parsec''
problem (e.g., \citealt{Escala+05b, Mayer+07, Callegari+09, Colpi+09,
  Hayasaki09}; see, however, \citealt{Lodato+09}), we expect the PTA
host galaxy to be the product of a relatively recent merger.  A
natural question to ask is whether such galaxies typically lie in the
field, or in the centre of a cluster.  We can answer this question
qualitatively by considering the dependence of the major merger rates
of the most massive dark matter halos on their environments.  Analyses
by \cite{Fakhma09} and \cite{Bonoli+10} of the Millennium simulation
results \citep{Springel+05} indicate that while the rate of major
mergers is enhanced in over-dense environments, this effect is weak:
for halo masses and redshifts of interest ($M\gta 10^{13}\Msol$ and at
$z\ltsim 1.5$), the ratio of merger rates between the most and least
over-dense regions is of order unity.  We interpret this result to
mean that there is no strong reason to search for PTA sources in
galaxy clusters as opposed to those in the field.

\subsubsection{The most massive halos}

One conservative way to estimate the number $N_g$ of host galaxy
candidates in the error box is to simply count the dark matter halos
that are massive enough to plausibly harbour the source SMBH binary.
We use the observational results of \cite{Dutton+10}, who infer a
double-power-law fit for the relation between the SMBH mass $M$ and
the host halo mass $M_{\rm halo}$ for local elliptical galaxies.  We
extrapolate their results to higher redshifts by postulating the
canonical $z$-dependence based on the theory for spherically
collapsing halos \citep[see, e.g.,][]{WL03},
\beq 
M_{\rm halo}(M,z)\propto F(z)\equiv\sqrt{
  \frac{d(z)}{d(0)}\frac{\Delta_{\rm c}(0)}{\Delta_{\rm c}(z)} }, 
\eeq
where $d(z)=-[(\Omega_{\rm m}/\Omega_{\Lambda})(1+z)^{3}+1]^{-1}$ and
$\Delta_{\rm c}(z)=18\pi^{2}+82d(z)-39 d^{2}(z)$.  We obtain 
\beq
M_{\rm halo}\approx 2.3\times 10^{13}M_{9}^{0.75} \left[\frac{1}{2}+13
  M_{9}^{1.77}\right]^{0.50} ~F(z)~\Msol ,
\label{eq:MM}
\eeq
where $M_{9}\equiv M/(10^{9}\Msol)$.  We estimate the number of
candidate host halos inside the three-dimensional PTA error box by
integrating the halo mass function of \citeauthor{Jenkins+01}
(\citeyear{Jenkins+01}; their equation 9) above $M_{\rm halo}$.

The most massive halos with $M_{\rm halo}\gta {\rm
  few}\times10^{14}\Msol$, which are associated with galaxy clusters,
may be expected to contain more than one plausible host galaxy.  Since
the halo mass function at $M_{\rm halo}\gta {\rm
  few}\times10^{14}\Msol$ drops much more steeply than linear with
mass, whereas the sub-halo mass function increases less steeply than
linear \citep{Giocoli+10},
most galaxies with halos masses $\sim 10^{13}\Msol$ will reside in the
field, rather than in groups and clusters.  The multiple occupancy of
massive galaxies in the most massive halos will then represent only a
small increase in our total counts of interlopers.  As the purpose of
the exercise in this section is to give order-of-magnitude estimates
for interlopers, we will neglect sub-halos in our analysis.

\subsubsection{The brightest galaxies}

A second way to estimate the number of candidate host galaxies is
through the $M-L$ relation, where $L$ is the luminosity of the host
galaxy.  Of particular interest is the fact that the $M-L$ relation
and the $M-\sigma$ relations are discrepant at the high-mass end (here
$\sigma$ denotes the velocity dispersion of the host).  The former
predicts higher masses for the most massive SMBHs, and higher number
densities for fixed BH mass \citep[][and references
within]{Lauer+07}. This therefore results in a greater number of
individually resolvable PTA sources \citep{SVV09}.

Since $\sigma$ is used to infer $M_{\rm halo}$, we expect that for a
fixed SMBH mass, the number of expected interloping host galaxies,
inferred from the $M-L$ relation, would also be greater than the
number of halos, inferred from the $M-\sigma$ and $\sigma-M_{\rm
  halo}$ relations.

To evaluate this different estimate quantitatively, we adopt the $M-L$
relation found by \citep{Lauer+07} for the most luminous core galaxies
in their sample,
\beq
M_{V}\approx -22.0 - 1.8\log_{10}M_{9},
\label{eq:ML}
\eeq
where $M_{V}$ is the $V$-band magnitude of the host galaxy.

To compute the number of sufficiently luminous galaxies, we use the
results of \cite{Gabasch+04, Gabasch+06}, who measured the luminosity
function in multiple wavelength bands between $150-900$ nm, and
studied the redshift evolution in each band out to $z\gta 2$.  The
luminosity function is given in the form of a standard Schechter
function \citep{Schechter76},
\begin{eqnarray}
\qquad\qquad\phi(M_{V})&=&\frac{2}{5}(\ln 10) \phi^{*}\left[10^{(2/5)(M_{V}^{*}-M_{V})}\right]^{\alpha_{0}+1}\nonumber\\
&&\qquad\times\exp\left[-10^{(2/5)(M_{V}^{*}-M_{V})}\right].
\end{eqnarray}
They set a constant value for the parameter $\alpha_{0}$ while fitting
$M^{*}_{V}$ and $\phi_{*}$ to a power-law redshift dependence of the
form
\begin{eqnarray}
\qquad\qquad M_{V}^{*}(z)&=&M_{V,0}^{*} + A \ln (1+z),\\
\qquad\qquad\phi^{*}(z)&=&\phi_{0}^{*} (1+z)^{B}.
\end{eqnarray}
The five fitting parameters $(\alpha_{0},M_{V,0}^{*},\phi_{0}^{*}
,A,B)$ vary with the wavelength band of the luminosity function.
Because the \citeauthor{Gabasch+04} results do not have fits for the
$V$-band, we interpolate the parameters between neighbouring
bands to obtain the following values: $\alpha_{0}\approx -1.3$,
$M_{V,0}^{*}\approx -21.1$, $\phi_{0}^{*}\approx 6.2\times
10^{-3}\Mpc^{-3}M_{V}^{-1}$, $A\approx -1.18$, and $B\approx -1.05$.

\subsubsection{The brightest AGN}
\label{subsec:AGNint}

Finally, a third method to identify plausible hosts is to search for
AGN that are luminous enough to be plausibly powered by a $M\sim
10^{9}\Msol$ SMBH.  
AGN activity is an ideal scenario for identifying the EM counterparts
of PTA sources,
as the interaction between a compact SMBH binary
and its accretion flow provide a natural physical mechanism
for eliciting a smoking-gun EM signature.
However, a significant uncertainty with this approach is
whether the host of a resolved PTA source is likely to be undergoing
an observable AGN episode.  While multiple studies have suggested that
galaxy mergers trigger AGN activity (refs. in \S\ref{sec:intro})
whether the two phenomena are causally related remains an open
question.  

Recently, \cite{Schawinski+11} suggested that the low Sersic indices
in most X-ray-selected AGN hosts at $1.5<z<3$ indicate that they are
disc galaxies, and therefore unrelated to mergers \citep[see,
however,][who suggest mergers can result in disc
galaxies]{Governato+09}.  Further, even if one accepts that there
exists a direct causal connection between galaxy mergers and luminous
AGN activity, it is uncertain whether such a trend extends to the most
massive galaxies at $z<1.5$.  The mass fraction of cold gas in massive
galaxies tend to decrease toward lower redshift, and gas-poor ``dry''
mergers are thought to play an important (if not dominant) role in the
assembly of giant elliptical galaxies at $z<1$, in the field as well
as in clusters \citep{vanDok05, Lin+08, Lin+10}.  On the other hand,
the amount of gas required to fuel a luminous AGN episode is a small
fraction of the total gas content of even very gas-poor galaxies; the
most luminous known AGN are situated in giant elliptical galaxies; a
plurality of mergers of massive galaxies at $z<1$ are gas-rich
\citep{Lin+08}; many early-type galaxies identified as undergoing a
dry merger have been found to contain detectable amounts of gas in
followup HI observations \citep[e.g.,][]{Donovan+07, S-B+09}; and even
though the hosts of the most luminous quasars tend to be ellipticals,
they are not exclusively so in the SMBH mass regime of interest ($M\ga
10^{8}\Msol$; e.g., \citealt{Percival+01, Floyd+04, Zakamska+06}, and
references therein).
We conclude that PTA sources powering luminous AGN activity
is a plausible scenario, and not merely an expedient assumption.

We parameterize the minimum luminosity
for the AGN counterpart in terms of the Eddington luminosity
$L_{\rm Edd}(M)=4\pi G M~ \mu_e m_{\rm p} c/\sigma_{\rm T}$,
where $\mu_e$ is the mean molecular weight per electron and
$\sigma_{\rm T}$ is the Thomson cross section:
\beq L_{\rm min}(M)=f_{\rm min}L_{\rm Edd}\left(M\right).
\label{eq:Lmin}
\eeq
We choose $f_{\rm min}=10^{-2}$ for our minimum Eddington ratio
 $L/L_{\rm Edd}$, motivated by the fact that this quantity
is observed to peak at $L/L_{\rm Edd}\sim 0.1-0.3$ \citep[e.g.,][]{Kollmeier+06}.

In order to estimate the number of AGN that are bright enough to
correspond to a PTA source with mass $M$, we adopt the
observationally motivated fits to the AGN luminosity function
given by \citeauthor{Hopkins+07b} (\citeyear{Hopkins+07b};
their equations 6, 18 and 20, and Table 3).

Note that although the Eddington ratio distribution
and the luminosity function cited above are expressed in terms
of the bolometric luminosity, they are actually
proxies for the optical luminosity.
\cite{Hopkins+07b} noted that their bolometric luminosity function
is effectively equivalent to the optical luminosity function,
and \cite{Kollmeier+06} uses the flux at 510 nm to estimate the
bolometric luminosity.
Our AGN interlopers are therefore optically luminous AGN,
and we assume nothing a priori about the X-ray and UV
emission of accreting PTA sources.
We will discuss the importance of searching for the EM
counterpart at optical wavelengths in \S \ref{subsec:emission}.

\subsection{Expected Counts of Interloping Galaxies}

If the individual pulsar contributions to the signal cannot be extracted,
as in the \citetalias{SesVec10} scenario, then the
chirp mass and luminosity distance of the resolved
source cannot be independently known,
and the source can only be localized within
a solid angle $\Delta\Omega$.
The only constraints on $\Mch$ and $D_{L}$
are then model-dependent, and come from
theoretical expectations for the population of
resolvable sources, given the detection
threshold of the array.
The upper end of the chirp mass distribution of
SMBH binaries,
along with the detector sensitivity,
sets a maximum luminosity distance $D_{L,{\rm max}}$
(equivalently, $z_{\rm max}$).
Similarly, the chirp mass distribution in the local
Universe determines a minimum chirp mass $\Mch_{\rm min}$
required for a PTA source to be resolved.
Note that since $M=\eta^{-3/5}\Mch\ge 2^{6/5}\Mch$,
the quantity $\Mch_{\rm min}$ also sets a
lower limit $M_{\rm min}\approx2.3 \Mch_{\rm min}$
on the gravitational mass.

The number of interloping halos can be expressed as
\beq
N_{\rm halo}^{(\rm SV)}=\frac{\Delta \Omega}{4\pi}
\int_{0}^{z_{\rm max}}\int_{M_{\rm halo}(M_{\rm min}, z)}^{\infty}
\frac{dn_{\rm halo}}{dM_{\rm halo}}~dM_{\rm halo}~
\frac{dV}{dz}~dz,
\label{eq:NhSV}
\eeq
where $n$ is the comoving number density of dark matter halos, and
$dV/dz=4\pi D_{L}^{2}~ dD_{L}/dz$ is the comoving volume element.  The
lower limit of the integral over halo mass is given by equation
\ref{eq:MM}.
Similarly, the numbers of interloping galaxies and AGN are given by
the expressions
\begin{eqnarray}
N_{\rm gal}^{(\rm SV)}&=&\frac{\Delta \Omega}{4\pi}
\int_{0}^{z_{\rm max}}\int_{-\infty}^{M_{V}(M_{\rm min},z)}\phi ~dM_{V}~
\frac{dV}{dz}~dz,
\label{eq:NgSV}\\
N_{\rm AGN}^{(\rm SV)}&=&\frac{\Delta \Omega}{4\pi}
\int_{0}^{z_{\rm max}}\int_{L_{\rm min}(M_{\rm min})}^{\infty}\frac{dn_{\rm AGN}}{dL} ~dL~
\frac{dV}{dz}~dz.
\label{eq:NASV}
\end{eqnarray}
The limits of integration over luminosity are taken from equations
\ref{eq:ML} and \ref{eq:Lmin}.
\begin{figure}
\centering
\epsfig{file=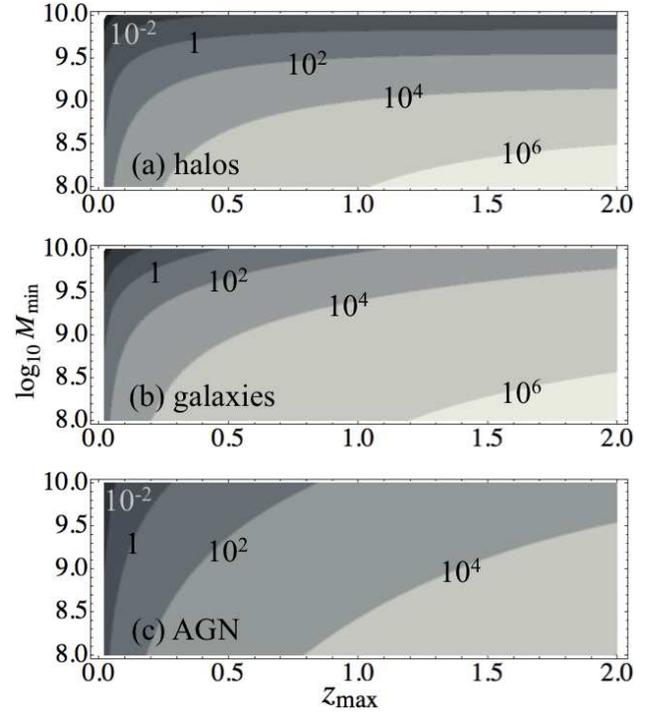, width=3.25in}
\caption{Estimates of the number of interloping host objects --- (a)
  massive dark matter halos, (b) luminous galaxies, and (c) luminous
  AGN --- in the conical error volumes suggested by
  \citetalias{SesVec10}.  The extent of the error volume is limited by
  $z_{\rm max}$, the maximum redshift at which PTAs can resolve an
  individual source, and the angular localization
  $\Delta\Omega=40\deg^{2}$.  The number of interlopers is calculated
  by assuming a minimum SMBH mass $M_{\rm min}$, which then sets the
  minimum host mass/luminosity through equations \ref{eq:MM},
  \ref{eq:ML}, and \ref{eq:Lmin}.  }
\label{fig:SVcount}
\end{figure}

\begin{figure}
\centering
\epsfig{file=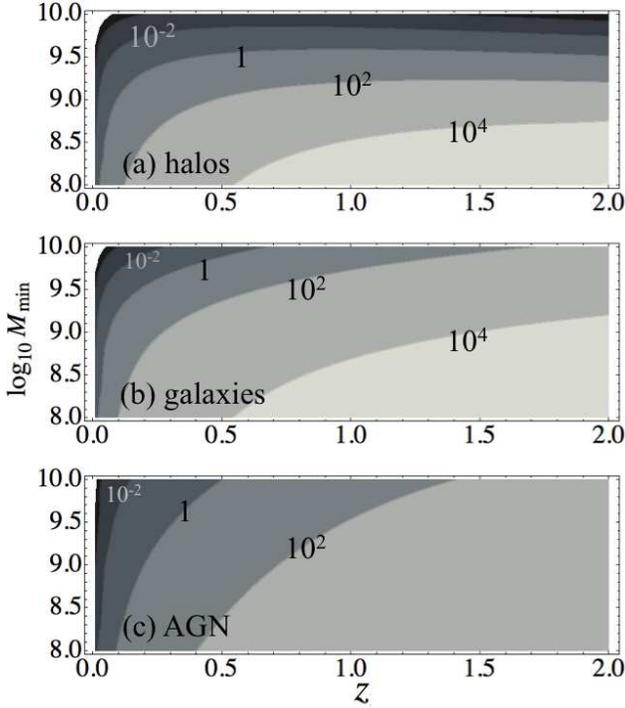, width=3.25in}
\caption{
Same as Figure \ref{fig:SVcount},
except that the error volume is calculated from the results
of \citetalias{CorCor10}, who assumed that the pulsar term
of the GW signal can be used to infer the luminosity distance
to the source binary. The error box is limited by the uncertainty
$\Delta D_{L}/D_{L}=20\%$ in the luminosity distance to the source,
and the angular localization $\Delta\Omega=3\deg^{2}$.
Note that whereas the horizontal axis
in Figure \ref{fig:SVcount} showed the maximal PTA detection range
$z_{\rm max}$, here it denotes the actual redshift $z$
of the source.
}
\label{fig:CCcount}
\end{figure}

We show in Figure \ref{fig:SVcount} the estimated number of
interlopers for the worst-case error box in the \citetalias{SesVec10}
scenario, assuming $\Delta\Omega=40\deg^{2}$, as a function of the
maximum redshift $z_{\max}$ and minimum BH binary mass $M_{\rm min}$.
Panels (a), (b) and (c) show the isonumber contours of the expected
number of interloping massive halos, luminous galaxies and luminous
AGN, respectively.  All three methods to estimate the number of
interlopers yield on the order of $N_g\ga 10^{2}$ for PTA sources
with $M>10^{9}\Msol$, if the redshift range is restricted to $z_{\rm
  max}\sim 1$.

The difference in the number
of interlopers between the top two panels (halos vs. galaxies) for the
most massive SMBHs arises because the observed SMBH samples yield an
internally inconsistent set of $M-\sigma$, $L-\sigma$ and $M-L$
relations, as mentioned above.  While the interpretation of this
inconsistency is beyond the scope of our paper, we note that
\cite{Tundo+07} discussed this issue, and concluded that the intrinsic scatter in the
relations produces a selection bias: using the observed BH samples
yields a biased $L-\sigma$ relation (too low $L$ for given $\sigma$).
This suggests that the $M-L$ galaxy relation we adopted may also be
biased and it under-predicts $L$; correcting this bias would decrease
the number of galaxy interlopers.

If the GW signal can be used to constrain $\mathcal{M}$ and $D_{L}$ of
the source via statistical inference, as suggested by
\citetalias{CorCor10}, then the numbers of interloping halos, luminous
galaxies and AGN are given by
\begin{eqnarray}
N^{\rm (CC)}_{\rm halo}(z)
&=&\frac{\Delta \Omega}{4\pi}\int_{z_{-}}^{z_{+}}\int_{M_{\rm halo}(M_{\rm min},z)}^{\infty}\frac{dn_{\rm halo}}{dM_{\rm halo}} ~dM_{\rm halo},
\label{eq:NhCC}\\
N^{\rm (CC)}_{\rm gal}(z)
&=&\frac{\Delta \Omega}{4\pi}
\int_{z_{-}}^{z_{+}}\int_{-\infty}^{M_{V}(M_{\rm min},z)}\phi ~dM_{V}~
\frac{dV}{dz}~dz,
\label{eq:NgCC}\\
N^{(\rm CC)}_{\rm AGN}(z)&=&\frac{\Delta \Omega}{4\pi}
\int_{z_{-}}^{z_{+}}\int_{L_{\rm min}(M_{\rm min})}^{\infty}\frac{dn_{\rm AGN}}{dL} ~dL,\label{eq:NACC}
\end{eqnarray}
respectively.  Above, the redshifts $z_{\pm}= z(D_{L}\pm\Delta D_{L})$
bound the radial extent of the error box.  We adopt $\Delta \Omega =
3\deg^{2}$ and $\Delta D_{L}/D_{L}=20\%$.  We ignore errors due to
weak lensing, which are expected to be on the order of several percent
for sources with $z\ltsim 1.5$ \citep{Kocsis+06, Hirata+10, ShangHaiman11}.
We do not place an upper limit on the host halo mass (or on the host
galaxy luminosity).  In principle, such an upper limit
could be computed, given PTA's observational error on the chirp mass
and the spread in the ratio between the chirp mass and the gravitational mass
of the binary (i.e., from the model-dependent mass ratio distribution
of resolved sources). 
For example, \citetalias{CorCor10} provide a chirp mass error estimate
of $\Delta\Mch\sim 5\%$. Converting the chirp mass to the gravitational
mass, however, can introduce a large uncertainty,
e.g. a factor of $\sim 2$ depending on whether
the mass ratio is $0.1$ or $1$.
Since the number density of interlopers
decrease rapidly with increasing halo mass (or luminosity), this
simplification should not affect our estimates.

In Figure \ref{fig:CCcount}, we plot the number of interloping host
candidates against the source redshift $z$.  Not surprisingly, the
prospects for EM identification improve dramatically in the
\citetalias{CorCor10} scenario.  For massive ($M\gta 10^{9}\Msol$)
resolved PTA sources, we anticipate that the error box will contain a
single host candidate at $z\ltsim 0.2$, and several hundred at
$z\ltsim 0.7$.  We expect only a single group-sized halo ($M\ga {\rm
  few}\times 10^{13}\Msol$) in the error box at any redshift in the
\citetalias{CorCor10} scenario.  Note that the number $N_g$ of
interlopers is not necessarily a monotonically increasing function of
$z$, as the decline in the number densities of the interloping objects
(in particular massive halos) competes with the increase in the
comoving size of the error boxes.

Our simple calculations show that in the scenario of
\citetalias{CorCor10}, resolved PTA sources with $M\ga 10^{9}\Msol$
and $z\ltsim 0.5$ are likely to have at worst dozens of interlopers in
the error box.  With this low number, one could conceivably perform
follow-up observations of each individual candidate.  If, on the other
hand, luminosity distances to the source cannot be determined, this
number increases to $\sim 10^{3}$, suggesting that it will become
extremely difficult to electromagnetically identify the source in the
absence of an obvious, tell-tale EM signature.

In practice, the number of interloping galaxies may be somewhat larger
than the value computed by equations
\ref{eq:NhSV}$-$\ref{eq:NACC}.  The halo mass of any given
candidate host system will not be known a priori, and the intrinsic
scatter in the $M_{\rm SMBH}-M_{\rm halo}$ ($M_{\rm SMBH}-\sigma_{\rm
  host}$) relation will lower the minimum halo mass threshold for
candidacy.  On the other hand, the simple calculations presented here
do not consider detailed demographic properties of resolved PTA
sources and plausible hosts, such as the presence of a nuclear stellar
core \citep[]{Makino97, Ravind+02, Milos+02, VMH03} or galaxy
morphology.
Including such factors in the analysis will narrow the
field of candidate hosts.

As we argue in \ref{subsec:emission},
candidate AGN counterparts may be further
vetted by examining their UV and X-ray
emission for features indicative of
a central SMBH binary
(see also \citealt{Sesana+11} for an in-depth
discussion of possible high-energy signatures
for pre-decoupling --- i.e., $t_{\rm GW}>t_{\nu}$ --- PTA sources).
In addition, PTA sources are sufficiently nearby that
it should be possible to observe an interloping AGN
together with its host galaxy.
It should therefore be possible to combine
the AGN emission, the galaxy luminosity
and the inferred SMBH mass to cross-check
candidate counterparts.

\section{Accretion Discs Around PTA-source Binaries}
\label{sec:discs}

Motivated by the results of the previous section that the number of
plausible host galaxies in the PTA error box may be tractable for
follow-up EM searches, we next model the EM emission properties of
SMBH binaries detectable by PTAs.  We focus our attention on SMBH
binaries that are undergoing luminous accretion, as these are the most
promising class of objects for EM identification.

Normalizing the binary mass $M$ and rest-frame period $P$ to the
typical orders of magnitude expected of resolved PTA sources,
$M=10^{9}\Msol M_{9}$ and $P=1\yr~P_{1}$, we write the semi-major axis
for the source binary as
\begin{eqnarray}
\qquad a(M,P)&=&101 ~M_{9}^{-2/3}P_{1}^{2/3}\frac{GM}{c^{2}}\nonumber\\
&=&2.23 \times 10^{-2}M_{9}^{1/3}P_{1}^{2/3}\pc.
\label{eq:aPTA}
\end{eqnarray}

Binaries detectable by PTAs have long overcome the so-called ``final
parsec'' problem.  The rest-frame time to merger for a binary with
mass $M$ and semi-major axis $a$, driven by GW emission alone, is
\begin{eqnarray}
\qquad t_{\rm merge}&\le&\frac{5}{256}\frac{c^{5}}{G^{3}M^{3}}\frac{a^{4}}{\eta}\nonumber\\
&=&1.9\times 10^{3} M_{9}\eta_{1:4}^{-1}\left(\frac{a}{10^{2}GM/c^{2}}\right)^{4}\yr \nonumber\\
&=&2.0\times 10^{3} M_{9}^{-5/3}\eta_{1:4}^{-1}P_{1}^{8/3}\yr
\label{eq:tmerge}
\end{eqnarray}
\citep{Peters64}.  Because typical resolved sources have
$M/\Mch=\eta^{-3/5}\sim 3$, we normalize the symmetric mass ratio
$\eta\equiv (M_{2}/M_{1})/[1+M_{2}/M_{1}]^{2}$ to the value
$\eta_{1:4}\equiv\eta(M_{2}/M_{1}=0.25)=0.16$.  Note that our ad hoc
translation between $M$ and $\Mch$ is not very sensitive to the value
of $q$; the ratio $M/\Mch$ varies by less than a factor of two in the
range $0.1\le M_{2}/M_{1} \le 1$.  The upper bound in equation
\ref{eq:tmerge} corresponds to binaries in circular orbits, with
eccentric orbits merging faster.
Recent work has shown that binaries may have eccentricities
as high as $\sim 0.6$ at decoupling (\citealt{Roedig+11};
see also \citealt{AN05, Cuadra+09}).
Thus, typical PTA-resolved sources
will coalesce on scales of $\sim 10^{3}$ years.  However,
exceptionally compact sources will coalesce on scales of several
years; 
for example, a binary with $P=0.1\yr$ 
--- approximately the lowest binary period that 
is expected to be observable with PTAs ---
will merge in $t_{\rm merge}\sim 4\yr$.

The tidal torques of the compact SMBH binary provide a particularly
promising mechanism for producing a tell-tale observable feature.
Theoretical calculations \citep{GT80, Artymo+91, ArtLub94, AN02,
  Bate+03, Hayasaki+07,MM08, Cuadra+09, Chang+10} robustly predict
that in geometrically thin circumbinary accretion discs, binary
torques can open an annular, low-density gap around the orbit of the
secondary.  The gas inside the gap accretes onto the individual SMBHs
while the gas outside is pushed outward by
the tidal torques.  The binary's tidal torques transfer orbital
angular momentum into the outer disc, causing the binary's orbit to
shrink gradually while maintaining a roughly axisymmetric circumbinary
gap.

The gap opens near the resonance radius $R\approx 3^{2/3}a \approx 2.08 a$
\citep{Artymo+91}.
Numerical simulations of thin circumbinary discs (see refs. in above paragraph)
produce gaps with an azimuthally averaged radius of $1.5-3$ times
the binary semimajor axis.
The exact size and shape of the gap is not easily characterized;
the geometry depends on the binary masses and orbital eccentricity,
as well as the efficiency of angular momentum transport within the disc.
Following \cite{MP05}, we parametrize the size of the gap as
$R_{\lambda}\equiv 2\lambda a$, where $\lambda\sim 1$ is a
dimensionless parameter.
We are interested in circumbinary discs that
are truncated inside $R_{\lambda}\sim 200 M_{9}^{-2/3}P_{1}^{2/3}
GM/c^{2}$ (equation \ref{eq:aPTA}).
Below, we model surface density profiles and thermal
emission spectra of such discs, as well as the thermal emission due to
leakage of gas into the cavity and onto individual SMBHs.

\subsection{Disk Properties and Binary Decay}

\subsubsection{Disk around a solitary SMBH}

Adopting a geometrically thin, thermal gray-body disc model
\citep[e.g.,][]{Blaes04,MP05}, we estimate the properties of
circumbinary discs around resolved PTA sources.  As a reference model,
let us consider a disc around a solitary SMBH.

Until recently, the standard $\alpha$-viscosity
prescription for accretion discs, in which the kinematic viscosity
scales proportionally with the total pressure in the fluid
(i.e., $\nu\propto \alpha (p_{\rm gas}+p_{\rm rad})$),
was thought to be thermally and viscously unstable \citep{SS76, Pringle76}.
Although magnetohydrodynamic shearing box simulations
by \citealt{HKB09} have since suggested that $\alpha$-discs
are actually thermally stable, they may still be viscously unstable \citep{LE74, HBK09}.
We therefore assume a viscosity prescription in which the kinematic
viscosity scales with gas pressure
(a.k.a. the ``$\beta$-disc'' model):
\beq \nu=\frac{2}{3}\frac{\alpha p_{\rm
    gas}}{\rho\Omega}=\frac{2}{3}\frac{\alpha k_{\rm B}T}{\mu m_{\rm
    p}\Omega}.
\label{eq:nu}
\eeq
This viscosity prescription is consistent with previous analyses
of thin circumbinary discs \citep{MP05, TM10}.

If discs around PTA sources are instead described by
the standard $\alpha$ viscosity prescription,
then they would have somewhat higher temperatures than
what we calculate below using the $\beta$ prescription.
The higher viscosity in the $\alpha$ model would also
result in the gas being able to follow the decaying binary
to closer separations. Both effects would result in the thermal
spectrum being somewhat harder than we calculate below,
but we expect our qualitative results to hold as long as the disc
is able to remain geometrically thin.

Observations of accreting Galactic compact objects
\citep[][and refs. therein]{King+07} and blazars \citep{Xie+09}
suggest $\alpha\sim 0.1-0.4$,
while studies of AGN continuum variability place a lower limit of
$\alpha\ga 0.01$ \citep{Starling+04}.
Numerical simulations are consistent with $\alpha \ga 0.1$,
with lower reported values possibly
being due to small sizes of the simulation box \citep{Pessah+07}.
We choose $\alpha=0.3$ as the fiducial value, and write $\alpha_{0.3}=\alpha/0.3$.
We assume that shear viscosity is the dominant mechanism
of radial gas transport in the circumbinary disk.

The surface density $\Sigma$ and the mid-plane temperature $T$ of the
disc are obtained through the following equations:
\begin{eqnarray}
\qquad\qquad\qquad\Xi(\Omega, T_{\rm p})\sigma_{\rm SB}T_{\rm p}^{4}
&=&\frac{9}{8}\nu\Sigma\Omega^{2}\label{eq:nutherm}\\
T_{\rm p}^{4}&=&\frac{4}{3\tau}T^{4}\\
\tau&=&\theta \kappa \Sigma\\
\Sigma&=&\frac{\dot{M}}{3\pi\nu}\label{eq:sigmdot}.
\end{eqnarray}
Above, $\Xi$ is the deviation of the bolometric flux from blackbody
due to the photons being thermalized above the mid-plane \citep[see,
e.g.,][]{Blaes04}, $T_{\rm p}$ is the temperature of the
thermalization photosphere, and $\tau$ is the optical depth between
the mid-plane and the thermalization photosphere. The
quantity $\theta$ is a porosity factor that relates the surface
density to the optical depth. We set it to $0.2$ following
\cite{Turner04} and express our results in terms of
$\theta_{0.2}=\theta/0.2$; however, most of the disc properties
are not very sensitive to this parameter.
We will use the dimensionless
parameter $\dot{m}\equiv \dot{M}/\dot{M}_{\rm Edd}$ to describe the
local accretion rate in units of the Eddington rate, assuming a radiative
efficiency of 0.1, i.e. $L_{\rm Edd}=0.1\dot{M}_{\rm Edd} c^{2}$.  

The disc scale height $H$ is evaluated in the
usual way:
\beq
c_{\rm s}^{2}\equiv \frac{p}{\rho}= H^{2}\Omega^{2}+4\pi G\Sigma H,
\label{eq:cs}
\eeq
where $c_{\rm s}\equiv \sqrt{p/\rho}$ is the isothermal sound speed
calculated from the total pressure $p=p_{\rm gas}+p_{\rm rad}$, and
the volume density $\rho$ of the disc is given by $\rho=\Sigma/H$.
The second term on the right-hand side of equation \ref{eq:cs} is
due to the disc's self-gravity \citep[e.g.,][]{Paczynski78}.

In the regions of interest, the primary source of opacity is electron scattering,
and radiation pressure dominates over gas pressure.
In this regime, the gray-body factor $\Xi$ can be approximated as
$\Xi\approx 0.17 (\Omega \yr)^{1/2}[T_{\rm p}/(10^{4}\K)]^{-15/8}$ \citep{TM10},
and with a little algebra we obtain the surface density profile in the disc:
\begin{eqnarray}
\qquad\qquad\Sigma (R)&\approx& 1.3\times 10^{6}\g\cm^{-2}\left(\frac{R}{100GM/c^{2}}\right)^{-6/17}\nonumber\\
&&\times M_{9}^{16/85}\dot{m}^{36/85}\alpha_{0.3}^{-4/5}\theta_{0.2}^{-1/5}.
\label{eq:Sig1BH}
\end{eqnarray}
A steady-state disc far from the central object
satisfies  $\dot{M}=3\pi\nu\Sigma={\rm constant}$, and so we have
$\nu\propto \Sigma^{-1}\propto R^{6/17}$.

\subsubsection{Circumbinary discs around orbit-decaying binaries}

After a circumbinary gap is opened, the SMBH binary undergoes several
stages of orbital decay.  Let us briefly examine the different stages,
and the orbital evolution timescale (or residence time) $t_{\rm
  res}\equiv a/|da/dt|$ for each.
Our goal here is to describe the structure of a dense gaseous annulus,
extending at least a factor of few in radius, that is created around
the PTA source. The annulus results from inward migration of the
binary from larger radii in a more extended accretion disc.  For a
more thorough discussion of the orbital decay of SMBHB binaries,
through various physical regimes in a thin disc, see,
e.g. \cite{HKM09}.

We begin with disc-driven orbital decay, in which the binary's tidal
torques transfer its orbital angular momentum to the surrounding gas.
At large orbital separations, the mass of the gas at the edge of the
cavity far exceeds the mass of the secondary.  In this regime,
analogous to disc-dominated Type II migration for proto-planets, the
binary's orbital evolution is limited only by the rate at which the
nearby gas can transport away angular momentum, i.e.
\beq
t_{\rm res}^{\rm (disc)}=t_{\nu}(R_{\lambda})=\frac{2R_{\lambda}^{2}}{3\nu (R_{\lambda})}.
\eeq
The tidal torques prevent the gas from flowing inward of
$R_{\lambda}$, and so the region inside the gap is starved.  Any gas
that is initially present will be depleted on the local viscous
timescale \citep{Chang+10}.  In standard steady-state thin-disc models
the viscosity is an increasing function of radius, so this drainage
occurs on timescales shorter than that of the binary's orbital decay.

When the mass of the secondary becomes comparable to the local disc
mass, the orbital decay slows down with respect to the local viscous
time.  The gas piles up immediately outside the cavity, forming a
decretion region in which the viscous torque
$\mathcal{T}_{\nu}=3\pi\nu\Sigma \Omega R^{2}$ is nearly constant with
radius \citep{Pringle91}.  We apply the analytic model of \cite{IPP99}
to calculate the residency time for this secondary-dominated migration
stage:
\beq
t_{\rm res}^{\rm (sec)}=\frac{\eta M}{4\pi R_{\lambda}^{2}\Sigma(R_{\lambda})}t_{\nu}(R_{\lambda}).
\eeq
Note that there are two competing effects influencing $t_{\rm
  res}^{\rm (sec)}$: the decay slows down as the local disc mass
decreases with respect to the secondary, but this is mitigated to a
small extent by the fact that $\Sigma$ outside the cavity increases
due to pile-up.  The enhancement of $\Sigma$ relative to that of a
disc around a solitary SMBH of the same mass as the binary (equation
\ref{eq:Sig1BH}) has the functional form \citep{IPP99}
\beq
\frac
{\Sigma^{\rm (binary)}}
{\Sigma^{\rm (solitary)}}
=\left\{
1+A\left[1-\left(\frac{R_{\lambda}}{R_{\lambda}^{\rm(disc/sec)}}\right)^{1/2}\right]
\right\}^{B}\left(\frac{R}{R_{\lambda}}\right)^{-1/2}
\label{eq:pileup}
\eeq
in the neighbourhood $R\gta R_{\lambda}$.  Above,
$R_{\lambda}^{\rm(disc/sec)}$ is the radius of the cavity when the
transition from disc-dominated to secondary-dominated migration
occurs, i.e. when $\eta M=4\pi R_{\lambda}^{2}\Sigma(R_{\lambda})$.
For reasonable parameter values,
$R_{\lambda}^{\rm(disc/sec)}>10^{3}GM/c^{2}$.
The dimensionless quantities $A$ and $B$ in
equation \ref{eq:pileup} depend on the viscosity and mass profiles
of the disc \citep[see][for details]{IPP99}.
We typically find that $A\sim 4$ and $B\sim 0.2$
in our disc models;
i.e., the fractional surface density enhancement
during secondary-dominated migration
is no greater than $(1+A)^{B}\sim 1.4$.

At yet smaller separations, the binary's orbital evolution begins to
be driven by GW emission.  Since binaries of interest here are far
from merging, GW emission can be approximated by the leading term in
the Newtonian quadrupole.  For circular orbits, the residence time is
given by \cite{Peters64}
\beq
t_{\rm res}^{\rm (GW)}=4t_{\rm merge}=\frac{5}{64}\frac{c^{5}}{G^{3}M^{3}}\frac{a^{4}}{\eta }.
\eeq
As the binary's orbital decay accelerates due to GW emission, the
pileup caused by secondary-dominated migration spreads out.  Past the
point where $t_{\rm res}^{\rm (GW)}\approx t_{\nu}(R_{\lambda})$, the
binary begins to outrun the disc, as the decay timescale for $a$
becomes rapidly shorter than that on which the disc can viscously
spread.

Let us now discuss the gravitational stability of the disc, based on
the stability criteria of a radiation-pressure dominated fluid
summarized by \cite{Thompson08}.  If the radiative diffusion timescale
is much shorter than the dynamical timescale, then the radiation
pressure does not stabilize the fluid and gravitational fragmentation
occurs on the same length scales as it would in the absence of
radiation pressure.  If the radiative diffusion timescale is much
longer than the dynamical time, which we find to be the case for our
disc models, then radiation pressure acts to make the fluid more
Jeans-stable.\footnote{The gas is susceptible to an additional weak
  diffusive instability that grows on the Kelvin-Helmholtz timescale,
  $t_{\rm KH}\sim \kappa c_{s}^{2}/(\pi Gc)$.  We find that the
  viscous timescale is shorter than the Kelvin-Helmholtz timescale ---
  i.e., the diffusive instability is irrelevant --- in all but the
  outermost annulus of the radiation-dominated region of our fiducial
  circumbinary discs (e.g., at $R=400 GM/c^{2}$, $t_{\rm KH}\sim
  2\times 10^{6}\yr$ and $t_{\nu}\sim 10^{5}\yr$).  Even in the small
  region where the disc is formally unstable to the diffusive
  instability, it is plausible that local turbulence can quench its
  growth.  We therefore assume the diffusive instability is unimportant.}.
We invoke the Toomre criterion, and assume that the disc is
gravitationally stable when the dimensionless parameter
\beq
Q(R) \equiv \frac{c_{\rm s}\Omega}{\pi G\Sigma}
\label{eq:Q}
\eeq
is greater than unity.
Note that the key effect of radiation pressure in this context
is that the sound speed $c_{\rm s}$ is computed from the total pressure,
not just the gas pressure.

A counterintuitive result is that in radiation pressure-dominated
discs, increasing the surface density --- or equivalently, the accretion rate ---
at fixed radius and disc parameters ($\alpha$, $\theta$, etc.)
makes it {\it more} stable against fragmentation.
This is a significant point, because it allows for the existence
of a copious amount of hot, gravitationally stable
gas near the binary.
We demonstrate this behavior as follows.
Combining equations \ref{eq:cs} and \ref{eq:Q},
we may write in general $Q=4\sqrt{x^{2}-x}$,
where $x\equiv p/(4\pi G \Sigma^{2})\ge 1$ is the ratio
of the pressure in the disk to its self-gravity.
It follows directly that the stability criterion $Q>1$
is equivalent to the condition $x>(2+\sqrt{5})/4\approx 1.1$.
Some algebraic manipulation of the disc equations (\ref{eq:nutherm} $-$ \ref{eq:cs})
gives another general expression for the region dominated
by electron scattering and radiation pressure:
\beq
x=\frac{9\theta\kappa_{\rm es}}{8\pi cG}\Omega^{2}\nu\Xi^{-1}\propto \frac{\nu}{\Xi}.
\eeq
In standard $\alpha$ and $\beta$ discs, increasing the surface density
with all other properties held constant increases the midplane temperature. 
This raises the value of $\nu$, and the graybody factor $\Xi$ decreases or
remains constant (e.g., $\Xi\equiv 1$ for blackbody discs).
Thus, in radiation pressure-dominated discs
the parameters $x$ and $Q$ increase with an increase in $\Sigma$
or the accretion parameter $\dot{m}$.
Increasing the accretion rate heats and ``puffs up'' the
radiation pressure-dominated portion of the disc
more than it acts to enhance self-gravity.
The opposite is true in the gas pressure-dominated region,
where one obtains
$x\propto \Sigma^{-1/3}H^{-1}\kappa^{1/3}\Xi^{-1/3}$;
increasing the surface density (or accretion rate) in the outer disc
drives it closer to gravitational instability.

We find that the disc is Jeans-stable inside a radius
\begin{eqnarray}
\qquad R_{Q}&\equiv& R(Q=1)\nonumber\\
&=& 550\frac{GM}{c^{2}}\alpha_{0.3}^{34/165}\dot{m}^{62/165}M_{9}^{-16/55}\theta_{0.2}^{17/55}.
\label{eq:RQ}
\end{eqnarray}
Equation \ref{eq:RQ} is valid for radiation pressure-dominated
regions only.
As a function of the accretion rate $\dot{m}$, $R_{Q}$
has a minimum value, where it coincides with the radius where
$p_{\rm rad}=p_{\rm gas}$.

The gas density profile in the outer regions $R>R_{Q}$, where
classical thin-disc models predict $Q<1$, is uncertain. One
possibility that has been explored by \cite{SirkoGoodman03} and others
\citep{Thompson+05, Levin07, Lodato+09} is that feedback mechanisms
(such as nuclear fusion from stars that formed in the disc or their
supernovae) inject sufficient energy as to maintain marginal
gravitational stability with $Q\approx 1$ in the outer regions.
However, the profile of the outer disc is not central to this study,
as we are interested in radiation from the central regions of the
disc, where the presence of a compact binary is most likely to produce
characteristic features that may distinguish them from accretion discs
around single SMBHs.  To keep our analysis as simple as possible, we
simply neglect the thermal radiation of the disc outside $R_{Q}$.

What is the accretion rate in the disc?  Uncertainty regarding the
outer gas distribution notwithstanding, quasars are able to
efficiently supply SMBHs of mass $>10^{8}\Msol$ with enough fuel to
maintain luminosities of $0.1-1L_{\rm Edd}$ for periods of $10^{6-8}
\yr$.  That most quasars radiate at just under the Eddington limit
while few exceed $L_{\rm Edd}$
\citep[see, e.g., ][for quasar Eddington ratios at $z\ltsim 1$]{Shen+08}
suggests that their luminosities are limited by radiative feedback,
rather than by the availability of fuel.
If this is the case, then the surface density in a discs around
compact binaries can be significantly greater than in a disc around a solitary
SMBH of the same total mass.  This is not because of the mass
accumulation of gas outside the binary's orbit, but because
such a disc would have a much lower luminosity owing to its low-density
central cavity.  If binary torques inhibit gas from accessing the
centre of the potential, then this effectively reduces the radiative
efficiency of the system, i.e. $L_{\rm disc}\ll \dot{M}c^{2}$.  We
find that even at $\dot{m}\sim 10$, the locally viscously dissipated
flux in our discs does not provide sufficient radiation pressure to
unbind gas from the local gravitational field.\footnote{We remind the
  reader that $\dot{m}$ is defined with respect to the Eddington limit
  assuming a radiative efficiency of $\sim 0.1$.  Strictly speaking,
  our circumbinary discs are not super-Eddington, even when the
  parameter $\dot{m}$ exceeds unity.}

\begin{figure}
\centerline{\hbox{
\includegraphics[width=3.25in]{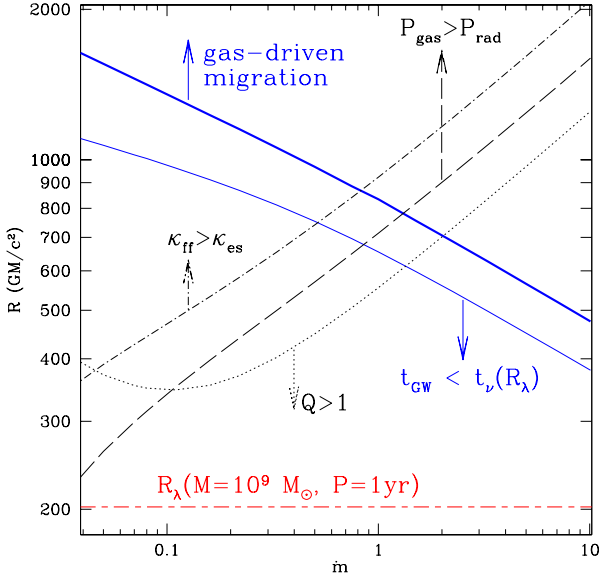}
}}
\caption[Transition radii for disc properties and binary migration]{
  In black lines, several transition boundaries within a steady-state,
  thin accretion disc are plotted as a function of the mass supply
  rate $\dot{m}=\dot{M}/(0.1L_{\rm Edd}/c^{2})$.  Binary parameters
  are $M=10^{9}\Msol$, $M_{2}/M_{2}=1/4$; disc parameters are
  $\alpha=0.3$, $\lambda=1$, $\theta=0.2$.  Plotted are the radii in
  the disc where the disc is marginally stable to gravitational
  fragmentation ($R_{Q}$; dotted lines); where the radiation pressure
  equals the gas pressure (dashed); and where free-free opacity equals
  electron-scattering opacity (dash-dot).  In blue lines, we plot the
  size of the circumbinary cavity $R_{\lambda}=2\lambda a$ where the
  binary's orbital decay transitions from being gas-driven to GW-driven
  (thick lines), and when the orbital decay timescale becomes shorter than
  the viscous timescale at the inner edge of the disc (thin lines).
  For reference, the radius of the cavity for a $10^{9}\Msol$ binary
  with a rest-frame period of $1\yr$ is plotted as a horizontal red
  dashed line.  }
\label{fig:radii}
\end{figure}

In Figure \ref{fig:radii}, we plot, for a circumbinary disc around a
binary with $M_{9}=1$, $M_{2}/M_{1}=1/4$, several transition radii as
a function of the mass accretion parameter $\dot{m}$.  We show the
radii at which the disc transitions from being radiation
pressure-dominated to gas pressure-dominated; from where the opacity
is dominated by electron scattering to free-free absorption; and from
Jeans-stable to unstable.  We note that in general, radiation pressure
acts to stabilize the disc against Jeans collapse, and that the radius
$R_{Q}$ closely corresponds to the radius where the disc becomes
radiation pressure-dominated.  We also plot the size of the cavity
$R_{\lambda}\sim 2a$ at which the binary's orbital evolution
transitions from being gas-driven to GW-driven, and the value of
$R_{\lambda}$ where $t_{\rm res}^{\rm GW}=t_{\nu}(R_{\lambda})$.  We find that
the disc is geometrically thin ($H/R\ll 1$) or marginally thin
($H/R\ltsim 1$) for $\dot{m}<10$.

In conclusion, Figure \ref{fig:radii} suggests that a Jeans-stable
circumbinary annulus could exist, instantaneously, around an
individually resolved PTA source, for any value of the supply rate,
extending at least by a factor of two in radius (from the inner radius
shown by the [red] dashed line, to the outer radius shown by the
[black] dotted curve).  However, in order for this annulus to be
created through the in-ward migration of the secondary BH from larger
radii in a more extended disc, we require that the disc is stable to
radii that extend beyond the gas/GW-driven transition.  This latter
requirement (i.e., the dotted [black] curve must lie above the thick
[blue] solid curve) means that in practice, the gaseous annulus exists
in PTA sources only if the mass supply rate is $\dot{m}\ga 1$.  As
argued above, while radiative feedback may disallow such high
(super-Eddington) rates in thin discs around a single BH, they can
naturally be maintained in discs around binaries, owing to their low
radiative efficiency.

\subsection{Surface Density Evolution of the Circumbinary Gas}
\label{sec:sigmaev}

Let us now address the surface density profile of the outer disc at
the time when the binary SMBH becomes observable by PTAs.

The tidal torque density $dT_{\rm tide}/dR$ is sharply peaked in a
narrow region that roughly coincides with the edge of the cavity
$R_{\lambda}$, preventing the gas from accreting inward.  Everywhere
else in the disc, the tidal torques are negligible compared to the
viscous torques.  The effect of the tidal torques in the region
$R\approx R_{\lambda}$ can thus be approximated as a boundary
condition prohibiting mass flow across $R_{\lambda}$ \citep{Pringle91,
  IPP99}:
\beq \dot{M}(R_{\lambda},t)=6\pi\nu\Sigma
\frac{\dd\ln(\Sigma\nu R^{1/2})}{\dd \ln R}\Big |_{R=R_{\lambda}}=0,
\label{eq:bc}
\eeq
Note that our disc is not steady-state, and the local mass flow rate
$\dot{M}$ need not be radially constant.

The surface density evolution of the circumbinary disc $R_{\lambda}$
is governed by the standard equation for viscous discs
\citep[e.g.,][]{Pringle81, FKR02} without including an explicit term
for the tidal torques:
\beq
\frac{\dd}{\dd t}\Sigma (R,t)=\frac{1}{R}\frac{\dd}{\dd R}
\left[ R^{1/2}\frac{\dd}{\dd R}\left(3\nu\Sigma R^{1/2}\right)\right].
\label{eq:diff}
\eeq

A semi-analytic solution for the thin-disc equation \ref{eq:diff}
with the boundary condition in equation \ref{eq:bc} was derived by
\cite{Tanaka11}, for a finite boundary $R_{\lambda}>0$ and a special
viscosity prescription $\nu\propto R^{n}$.  The solution can be
written in the form
\beq
\Sigma (R,t)=\int_{R_{\lambda}}^{\infty}G(R,R^{\prime}; t, R_{\lambda})
~\Sigma_{\rm init}(R^{\prime})~ dR^{\prime},
\label{eq:sigt}
\eeq
where $G(R,R^{\prime}; t, R_{\lambda})$ is the Green's function
specific to the boundary condition and the chosen value of the
viscosity power-law index $n$; and $\Sigma_{\rm init}(R)$ is an
arbitrary initial density profile.  We find that our discs satisfy
$\nu\propto R^{0.4}$ inside $R<10^{3}GM/c^{2}$, and thus adopt
$n=0.4$.

In order to model a thin accretion disc around a GW-driven SMBH
binary, we modify the solution of \cite{Tanaka11} in two ways.  First,
we derive a more general Green's function to allow for a boundary
condition with nonzero mass flux across the inner boundary:
\beq
\dot{M}(R_{\lambda},t)
=f_{\rm leak}\dot{M}_{\rm ss}(R_{\lambda},t).
\label{eq:bc2}
\eeq
Above, $\dot{M}_{\rm ss}=3\pi\nu\Sigma$ is the standard accretion rate
for expected of a steady-state disc, and $0\le f_{\rm leak}<1$ is a
numerical factor representing the incomplete suppression of gas inflow
into the cavity.  The case $f_{\rm leak}=0$ corresponds to total
suppression of accretion by the binary's tidal torques.

The boundary condition in equation \ref{eq:bc2} is motivated by
results from numerical simulations, which show that in general the
binary torques do not completely prevent accretion into the gap, but
rather allow some gas to leak into the centre of the disc with a
suppressed mass flux $f_{\rm leak}\ltsim 0.1$ \citep[e.g.,
][]{ArtLub96, Gunther+04, Ochi+05, MM08}.  We choose $f_{\rm
  leak}=0.1$ as our fiducial value.

The long-term behavior of the gas is to pile up near the cavity and
satisfy the power-law $\nu\Sigma \propto R^{(f_{\rm leak}-1)/2}$ in
the vicinity of the boundary.  In comparison, a steady-state around a
solitary central mass disc satisfies $\nu\Sigma={\rm constant}$, and a
boundary condition imposing zero inward mass flux satisfies
$\nu\Sigma\propto R^{-1/2}$ at the boundary.  Note, however, that the
value of $f_{\rm leak}$ does not have a strong effect on the mass
profile (and hence the luminosity produced) outside the cavity-opening
radius $R_{\lambda}$; the fractional surface density enhancement due
to secondary-dominated migration is typically of order unity.

The gas that enters the cavity does so in nearly radial orbits
\cite{MM08}, and so is dynamically decoupled from the circumbinary
disc.  Thus, the surface density and mass flux inside $R_{\lambda}$
can consistently be disregarded in the Green's function formalism.
The leaked gas can presumably form accretion discs around one or both
SMBHs \citep{Hayasaki+07}, presumably at the usual AGN radiative efficiency $\sim 0.1$.
The mass supply rate of such circum-secondary (or -primary) discs will be modulated by
$f_{\rm leak}\dot{M}_{\rm ss}$, which decreases
as the binary outruns the circumbinary gas.
Because the viscous time at the outer edge
of such discs are shorter than at $R_{\lambda}$, they will be nearly steady-state,
with the surface density profile at any given time being
determined by the instantaneous mass flux into the cavity.
Thus, the bolometric luminosities of the discs around
each disc may be roughly expressed as
$L< 0.1~f_{\rm leak}\dot{M}_{\rm ss}(R_{\lambda})c^{2}$,
and would be Eddington-limited by the potential of
the individual black holes they orbit. This suggests that if the
quantity $f_{\rm leak}\dot{m}$ exceeds unity, the region inside the
cavity would develop radiation-driven outflow winds.  Thus, the
parameter $f_{\rm leak}$ affects the energetic output due to accretion
inside the cavity far more than it does that of the circumbinary disc.

The Green's function for the boundary condition in equation \ref{eq:bc2}
is given by
\begin{eqnarray}
G(R,R^{\prime}; t, R_{\lambda})
&=&\left(1-\frac{n}{2}\right)R^{-n-1/4}R^{\prime 5/4}\nonumber\\
&&\hspace{-0.1in} \times\int_{0}^{\infty}
\left[J_{\ell}(ky)\tilde{Y}_{\ell}(ky_{\lambda})-Y_{\ell}(ky)\tilde{J}_{\ell}(ky_{\lambda})\right]
\nonumber\\
&&~\times \left[J_{\ell}(ky^{\prime})\tilde{Y}_{\ell}(ky_{\lambda})-Y_{\ell}(ky^{\prime})\tilde{J}_{\ell}(ky_{\lambda})\right]
\nonumber\\
&&~ \times\left[\tilde{J}_{\ell}^{2}(ky_{\lambda})+\tilde{Y}_{\ell}^{2}(ky_{\lambda})\right]^{-1}
\nonumber\\
&&~ \times\exp\left[-3\Lambda k^{2}t\right]~k~dk.
\end{eqnarray}
Above, $\ell=1/(4-2n)$ and $\Lambda= (1-n/2)^{2}\nu R^{-n}$ are
constants.  We have introduced the variables $y=R^{1-n/2}$,
$y^{\prime}= R^{\prime 1-n/2}$ and $y_{\lambda}= R_{\lambda}^{1-n/2}$,
as well as the functions
\begin{eqnarray}
\tilde{J}_{\ell}(x) &=&x ~J_{\ell-1}(x)-\frac{f_{\rm leak}}{2-n} J_{\ell}(x)\qquad{\rm and}\\
\tilde{Y}_{\ell}(x) &=&x ~Y_{\ell-1}(x)-\frac{f_{\rm leak}}{2-n} Y_{\ell}(x).
\end{eqnarray}
Taking $f_{\rm leak}\rightarrow 0$ leads to the solution given in
\citeauthor {Tanaka11} (\citeyear{Tanaka11}; his equation 42) imposing
$\dot{M}(R_{\lambda})=0$.

The second modification to the Green's function formalism is to allow
the inner boundary to move inward as an explicitly known function of
time, i.e., $R_{\lambda}(t)=2\lambda a(t)$.  This is done through a
time-weighted superposition of different Green's functions at
intermediate values of $R_{\lambda}(t)$.
The ``master'' Green's function $\mathcal{G}$
for a time-dependent boundary takes on the form
\beq
\mathcal{G} (R,R^{\prime}; t) = \int \frac{\dd G (R,R^{\prime}; t, R_{\lambda})}{\dd R_{\lambda}}\frac{d R_{\lambda}}{d t}~d t.
\eeq
A derivation and expression for $\mathcal{G}$ is
provided in the Appendix.

We take our initial condition as the disc profile when
$t_{\rm res}^{\rm GW}=t_{\nu}(R_{\lambda})$,
just as the binary is just beginning to outrun the circumbinary gas.
We note that the torques exerted by the
disc are not entirely negligible at this stage. For simplicity, we
approximate the contribution of the disc torques to the orbital decay
rate, $a/(da/dt)_{\rm disc}= t_{\rm res}^{\rm (sec)}$, as being
constant.  This is justified as follows.  The disc torques are a weak
function of $R_{\lambda}$, at least as long as the quantity
$\nu(R_{\lambda})\Sigma(R_{\lambda})$ is comparable to the
steady-state value.  Once GW emission dominates the orbital decay, the
contribution of disc torques becomes quickly negligible regardless of
the value of $\nu(R_{\lambda})\Sigma(R_{\lambda})$.

The orbital decay is then given by 
\beq
\frac{da}{dt}=\left(\frac{da}{dt}\right)_{\rm GW}+ \left(\frac{da}{dt}\right)_{\rm disc}
=\frac{a}{t_{\rm res}^{\rm(sec)}}+\frac{64}{5}\frac{c^{5}}{G^{3}M^{3}}\frac{\eta}{a^{3}}.
\eeq
With the assumption that $t_{\rm res}$ is roughly constant, this has
the analytic solution
\beq
t=\frac{t_{\rm res}^{\rm(sec)}}{4}\ln \left[
\frac{a_{0}^{4} + 64c^{5}\eta t_{\rm res}^{\rm(sec)}/(5 G^{3}M^{3})  }
{a^{4}(t) + 64c^{5}\eta t_{\rm res}^{\rm(sec)}/(5 G^{3}M^{3})  }
\right]
.
\eeq

\begin{figure}
\centerline{\hbox{
\includegraphics[width=3.25in]{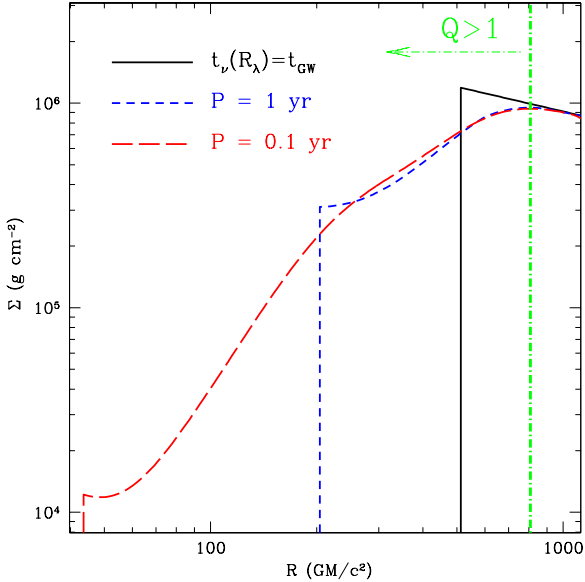}
}}
\caption[Surface density profile for a circumbinary disc around a PTA
source]{The surface density profiles $\Sigma$ for a circumbinary disc
  around a PTA source.  The binary's
  mass is $10^{9}\Msol$ and its mass ratio is $M_{2}/M_{1}=1/4$.  We
  adopt a moderately high value of the accretion parameter
  $\dot{m}=3$, and assume that the circumbinary gas
  can leak into the cavity at the rate given in equation \ref{eq:bc2}, with
  $f_{\rm leak}=0.1$ .  The solid black curve
  shows the surface density profile when GW emission begins to
  dominate the binary's orbital decay ($t_{\nu}(R_{\lambda})=t_{\rm
    GW}$, $R_{\lambda}= 510 GM/c^{2}$, $P=11\yr$).  Using the
  semi-analytic method described in the text, we solve for the surface
  density profile in the disc at later times, when $P=1\yr$
  (short-dashed blue line) and $P=0.1\yr$ (long-dashed red line).  The
  dot-dash green line denotes the radius inside which the circumbinary
  disc is stable against Jeans collapse.  }
\label{fig:Sigprof}
\end{figure}

In Figure \ref{fig:Sigprof}, we plot a model surface density profile
of the circumbinary disc around a binary with $M_{9}=1$ and
$M_{2}/M_{1}=1/4$.
As the initial condition, we take a steady-state
surface density profile
with $\dot{m}=3$,
at the time when $t_{\rm (GW)}=t_{\nu}(R_{\lambda})$
(solid black curve).\footnote{
Strictly speaking, the surface density profile at this time
should deviate somewhat from the steady-state one.
During secondary-dominated migration,
the circumbinary surface density profile $\Sigma(R\ga R_{\lambda})$
can become greater than the steady-state profile by at most a factor of $\sim 1.4$
(equation \ref{eq:pileup}; note that the pileup must be smaller
if one accounts for the fact that $f_{\rm leak}>0$).
Prior to our initial condition, GW emission accelerates
the binary's orbital evolution,
and the circumbinary pileup spreads out;
however, the surface density does not decrease
below the steady-state profile, since $t_{\rm res}>t_{\nu} (R_{\lambda})$.
A difference in $\Sigma$ of less than $40\%$
is insignificant compared to the other theoretical uncertainties,
and we employ the steady-state profile for simplicity.
}
Initially, the disc has a cavity
radius of $R_{\lambda}= 510 GM/c^{2}$ and a rest-frame period of
$P=11\yr$.
We then evolve the profile using our Green's function to
when the orbital period is $P=1\yr$ (short-dashed blue curve) and
$P=0.1\yr$ (long-dashed red curve).  We denote with a green dot-dashed
line the radius where $Q=1$, beyond which the disc is expected to be
susceptible to the Jeans instability.  All the disc profiles are
truncated at twice the binary's semi-major axis, and lose mass across
this radius through the boundary condition in equation \ref{eq:bc2},
with $f_{\rm leak}=0.1$. Note that the boundary radius $R_{\lambda}$
moves inward faster than the gas can pile up.  We see that a small
amount of gas is able to follow the binary's orbital decay, even
though the bulk of the circumbinary disc is getting left behind by the
inspiraling binary.

\subsection{Thermal Emission of Accreting PTA Sources}
\label{subsec:emission}

Since the innermost gas is missing from the accretion discs around PTA
binaries, it is probable that their accretion flows will emit less UV
and thermal X-rays compared to ordinary AGN powered by solitary BHs.
The question then is how UV- and X-ray-deficient these objects are;
the answer depends on how much gas is able to follow the binary's
decaying orbit, and how much of this gas is further able to leak into
the cavity and accrete onto individual SMBHs.  This is a complex
problem characterized by dynamical richness in 3D, and considerable
theoretical uncertainty of the underlying fluid physics.  With the
above caveat in mind, we will use as a first approximation the toy
surface density evolution model introduced in \S\ref{sec:sigmaev} to
estimate the thermal emission from an accreting PTA source.

\begin{figure}
\centerline{\hbox{
\includegraphics[width=3.25in]{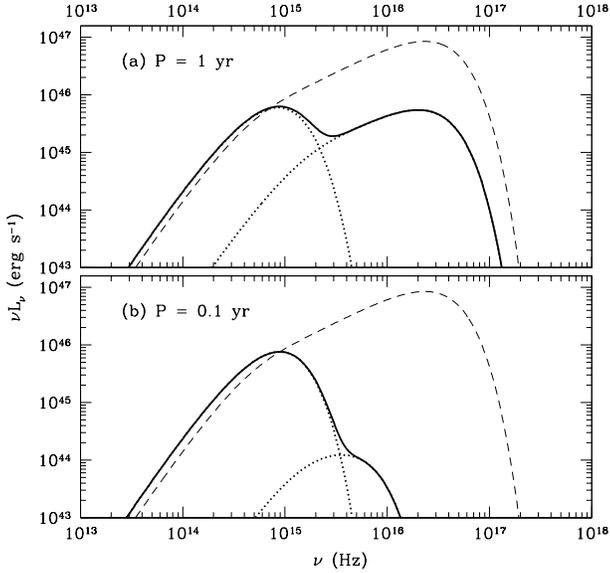}
}}
\caption[Model spectral energy distribution for an accreting PTA
source]{ We plot the spectral energy distribution from the thin,
  viscous circumbinary accretion disc from Figure \ref{fig:Sigprof}
  (red curve; $M=10^{9}\Msol$, $P=1\yr$), when the source has a period
  (a) $P=1\yr$ and (b) $P=0.1\yr$.  We assume that some of the gas at
  the cavity/disc boundary is able to accrete into the cavity at a
  suppressed rate and fuel a small accretion disc around the secondary
  SMBH (see text for details).  The spectra from individual discs are
  shown in dotted lines, with the circumbinary cavity emitting at
  lower frequencies.  The combined spectrum from the circumbinary and
  circum-secondary accretion discs is shown in solid thick lines.  We
  plot for reference the model AGN spectrum (thin dashed) for an
  Eddington-limited thin disc around a single SMBH with the same mass
  as the binary.  The spectra for the PTA source are UV and X-ray weak
  compared to a standard thin disc model around a single SMBH.
  However, the infrared and optical emission, mostly produced in the
  circumbinary disc, is similar to what would be expected for a
  single-SMBH disc.  }
\label{fig:spectra}
\end{figure}

In Figure \ref{fig:spectra}, we show the thermal spectrum for a
$M_{9}=1$, $M_{2}/M_{1}=1/4$ PTA source, calculated from the
circumbinary gas surface density profiles in Figure \ref{fig:Sigprof}.
We have plotted (i) the circumbinary disc with $\dot{m}=3$ inside the
radius where $Q=1$ (dotted, left hump); (ii) an accretion disc around
the secondary SMBH (dotted, right hump) fueled by leakage into the
cavity and truncated at the Hill radius, for which we use $R_{\rm
  H}\sim 0.5\eta ^{1/3}a$; and (iii) the combined emission of the two
discs (solid thick line).  For comparison, we also show the spectrum
for an Eddington-limited thin disc (dashed lines) around a solitary
SMBH with the same mass as the binary.  For simplicity, we have
assumed that all of the gas leaked into the cavity fuels a
circum-secondary disc.  We have plotted spectra when the source has a
binary orbital period of $P=1\yr$ and when $P=0.1\yr$.

The result is what would be expected intuitively.  The infrared and
optical flux, which is produced almost exclusively in the circumbinary
disc, does not vary greatly from what is expected from a standard thin
disc.  However, the flux drops precipitously below wavelengths of
$\lambda\ltsim 3000 \AA$ ($\nu>10^{15}\Hz$ in the figure).  This is in
stark contrast to most unobscured quasars thought to be powered by
$\sim 10^{8-9}\Msol$ SMBHs, which have their {\it brightest} emission
in the rest-frame near-UV near their Lyman-$\alpha$ line. The
bolometric luminosity of the accreting PTA source is
roughly $\sim 0.03L_{\rm Edd}$
(i.e., $L/L_{\rm Edd}\sim 10^{-2}\dot{m}$)
for $P=1\yr$,
and $\sim 10^{-3}L_{\rm Edd}$ for $P=0.1\yr$.
The optical and infrared emission is dominated
by the circumbinary disc,
whereas the UV and X-rays are produced by
circum-secondary accretion fueled by leakage of circumbinary
gas into the cavity.
As the binary evolves toward shorter periods,
the circum-secondary disc is depleted
--- the viscous time at the Hill radius is typically a few
hundred years, shorter than the time to binary merger ---
and as a result, less gas is able to leak
into the cavity, decreasing the UV and X-ray emission.
The degree to which the UV and X-ray emission is suppressed depends
on the model parameters (in particular $f_{\rm leak}$) and on the
binary period.
Note that the system may still be luminous in hard X-rays
due to inverse Compton scattering by
a coronal electron plasma \citep[][]{Sesana+11}.

We also note that the downturn in the near-UV flux
at $\lambda \ltsim 300~{\rm nm}$
could help distinguish PTA sources from single-SMBH AGN.
This feature will be observable in the optical if the source
redshift is high; e.g., at $z=1$ it will be in the {\it V} band.
Hence, even in the optical, this source will have an unusual color:
it will appear fainter in the {\it U} and {\it B} bands than a typical AGN.
The downturn could be distinguished from reddening due to
dust obscuration through the deviation from the power-law
spectral shape of dust reddening.

We propose that once an individually resolved PTA source is detected
and its error box determined, searching for AGN with weak UV emission
lines (e.g., Ly $\alpha$) and/or weak soft X-ray emission is a
promising method to narrow the field of interlopers.  AGN whose
soft X-ray fluxes are weaker by more than a factor of 10 compared to
the average have indeed been detected, and are estimated to constitute at most
$\sim 1\%$ of the general AGN population \citep[e.g.,][and
refs. therein]{Brandt+00, Leighly+07a, Gibson+08, Wu+11}.  There have
also been observations of quasars with exceptionally weak lines
\citep{Diamond+09}; these objects have infrared and optical emission
consistent with those of typical luminous AGN, and also tend to be X-ray
weak \citep{Shemmer+09}.
That X-ray weak AGN are so rare suggests that it will be possible
to narrow the number of interlopers in a typical PTA error box by a
factor of $\approx 100$, i.e. either to a handful of objects, or
yielding a unique EM counterpart candidate. It is possible, furthermore,
that some of these rare X-ray weak AGN are in fact the SMBH binaries
that PTAs will be detecting.

Our results also strongly suggest that AGN counterparts to PTA
sources should draw from optically selected surveys, as their nature
makes them likely to be missed by X-ray searches
(see, however, \citealt{Sesana+11}, who investigate the possible
X-ray searches of PTA source binaries that have not yet decoupled).
The {\it Large Synoptic Survey Telescope}\footnote{http://www.lsst.org/lsst} should
be able to detect all of the optically luminous AGN in the PTA
error box within $z\sim 1$.  It may be possible to follow up
candidates individually, but comparing the optical data with that of
wide-field X-ray surveys such as {\it
  MAXI}\footnote{http://maxi.riken.jp/top/}, {\it
  eROSITA}\footnote{http://www.mpe.mpg.de/erosita/} or {\it Wide Field
  X-ray Telescope}\footnote{http://www.wfxt.eu/home/Overview.html}
would greatly facilitate the multi-wavelength search for counterpart
candidates inside the error box.

Additional follow-up studies of candidates may further corroborate the
identification of a counterpart.  For example, the gas that leaks
radially into the cavity can shock-heat the outer edge of the
circum-secondary (or circum-primary) disc and produce hot spots.  The
viscously dissipated luminosity of a circum-secondary disc is roughly
$L_{\rm disc2}\ltsim (1/2)GM_{2}\dot{M}_{2}/R_{\rm ISCO,2}$, where
$\dot{M}_{2}\le f_{\rm leak}\dot{M}(R_{\lambda})$ is the mass supply
rate of the circum-secondary disc and $R_{\rm ISCO,2}$ is the radius
of innermost stable circular orbit around the secondary.  The
time-averaged power per unit mass of the hot spots is limited by the
amount of kinetic energy the flow can deposit at the outer edge of the
circum-secondary disc, i.e. $L_{\rm hot}\ltsim
GM_{2}\dot{M}_{2}/R_{H}$.  It follows directly that the time-averaged
ratio between between the hot-spots and the intrinsic luminosity of
the circum-secondary disc is \beq \frac{L_{\rm hot}}{L_{\rm disc2}}
\ltsim \frac{R_{\rm ISCO,2}}{R_{H}} \sim \eta^{-1/3}\frac{R_{\rm
    ISCO,2}}{a} \ga\left(\frac{a}{GM/c^{2}}\right)^{-1} .  \eeq In
other words, the time-averaged power of a hot spot is of order $\ga
1-10\%$ of the circum-secondary disc luminosity for resolved PTA
sources.  In principle, the luminosity of any single flare could be
much greater.  Because streaming into the cavity is expected to be
modulated quasi-periodically by the binary's orbital period
\citep[e.g.,][]{Hayasaki+07, MM08},
EM counterparts of resolved PTA sources may be characterized by periodic
UV flares.

In the same vein, if the orbital plane lies close to the line of
sight, the UV lines would display strong periodic Doppler shifting
with respect to the optical emission, modulated at the binary's
orbital period \citep[e.g.,][]{HalpFilipp88}.
Thus, monitoring
candidate counterparts for periodic or quasi-periodic variability on
orbital timescales may prove a fruitful route for identification
\cite{HKM09}.
As a proof of this concept, we note that \cite{BorLau09}
recently reported a candidate SMBH binary, with two sets of broad
emission lines separated by $3,500 {\rm km~s^{-1}}$, with inferred
component masses of $M_1=10^{8.9}~{\rm M_\odot}$ and
$M_2=10^{7.3}~{\rm M_\odot}$.  The binary interpretation, however,
could be ruled out by the lack of any change in the velocity offset
between two spectra taken $\approx 1$ year apart \citep{Chornock+09}.

Lastly, we consider the scenario of \cite{Chang+10},
in which the circum-primary disc brightens
prior to merger due to tidal excitation by the shrinking binary.
The power generated by this process can
be approximated as (see their equation 15)
\begin{eqnarray}
L_{\rm tide}&
\sim& \frac{GM M_{\rm in}}{2at_{\rm merge}}\nonumber\\
&\sim& 1.4\times 10^{43} 
M_{9}^{7/3}\eta_{1:4}P_{1}^{-10/3}\frac{M_{\rm in}}{100\Msol}
{\rm erg~s}^{-1},
\label{eq:tidalex}
\end{eqnarray}
where $M_{\rm in}$ is the mass of the circum-primary disc.
Extrapolation of the \cite{Chang+10} results
to binaries with mass $M\sim 10^{9}\Msol$
(their calculations only considered binaries up to $M= 10^{8}\Msol$)
suggests a value of $M_{\rm in}\sim 100\Msol$.
Similar values are obtained by estimating
the disc mass that can be fueled by
gas leaking into the cavity with $f_{\rm leak}\sim 0.1$,
when the time to merger is comparable to the
viscous timescale at the outer edge of
the circum-primary disc.

Equation \ref{eq:tidalex} suggests that the power
produced by tidal excitation of the circum-primary
disc is negligible compared to the thermal disc
emission if the binary period is $P\sim 1\yr$.
However, for sources with $P\sim 0.1\yr$,
the tidally excited emission would rival
the bolometric output of the thermal emission.
The tidal component, which would have a peak
frequency in the UV and soft X-rays,
will brighten dramatically prior to merger
on timescales of several years to decades.
Even though $P\sim 0.1\yr$ sources are
predicted to be much rarer and also much
more difficult to resolve individually with PTAs,
they present tantalizing possibilities for observing 
EM signatures that are directly related
to binary coalescence.

\section{Conclusions}

In this paper we considered the possibility that individually resolved
PTA sources --- SMBH binaries with $M\sim 10^{9}\Msol$,
$M_{2}/M_{1}\sim 1/4$, $P\sim 0.1-1 \yr$ and $z\ltsim 1.5$ --- may be
identified with EM observations if they reside in gas-rich
environments.  Multi-wavelength observations of such systems would
allow for studies of an AGN in a system that is known to harbour a
compact SMBH binary, thus providing a unique window into gas accretion
in a rapidly time-varying gravitational potential.  If, as suggested
by \cite{CorCor10}, PTAs can constrain the luminosity distance to
individually resolved sources, these SMBHs can be used as ``standard
sirens'' to measure the cosmic expansion history.  Interestingly, the
predicted redshift distribution of these sources lies between $0.1
\ltsim z \ltsim 1.5$, a range comparable to that of the deepest
Type-Ia supernovae surveys and where {\it LISA} detection rates
are expected to be low \citep{Sesana+07}.

Our findings can be summarized as follows:
\begin{itemize}

\item The number of interloping massive halos and AGN that may be
  confused with the PTA source is typically $N_g\sim 10^{4}$ for a
  $10^{9}\Msol$ binary if the contributions to the signal from
  individual pulsars are not identifiable in the PTA data.

\item In the more optimistic case, the pulsar term can be constrained
  and utilized to better determine the sky location of the source as
  well as constrain its redshift and mass. The number of interlopers
  can then drop to $N_g\sim 10-100$ at $z\sim 0.5$, and perhaps to
  $N<1$ for close sources at redshifts as close as $z\ltsim 0.2$.

\item By considering the orbital evolution history of an accreting PTA
  source, first by tidal interactions with the circumbinary gas and
  then by GW emission, we showed that a gaseous accretion disc around
  the source can be expected to be gas-poor both inside and
  immediately outside its orbit.  They would thus have optical and
  infrared luminosities comparable with typical quasars, while
  exhibiting low thermal X-ray luminosities and weak UV emission lines.
  The downturn in flux below $300~{\rm nm}$ could be discerned
  by optical observations if the source redshift is $z\ga 1$.
  The leakage of circumbinary gas into the cavity may shock
  the circumprimary (-secondary) disc and cause substantial quasiperiodic
  fluctuations in the UV and X-ray.
  Searching for AGN in the PTA error box with one or more of these
  atypical characteristics could lead to the identification of a single EM
  counterpart.  Further monitoring candidate counterparts for
  periodicity and other theoretically predicted pre-coalescence signatures
  may also aid identification.
\item 
  The above emission features are general for AGN powered
  by thin accretion discs around compact SMBH binaries.
  They could thus be used to discover such systems even in the
  absence of a GW detection. This is a particularly interesting
  possibility, as a small fraction ($\ltsim 1\%$) of optically luminous
  AGN is known to have low X-ray luminosities and unusually weak
  emission lines. Upcoming wide-field surveys in the optical
  and X-rays should together discover many more such AGN,
  as well as observe them over temporal separations
  of weeks to months.
  Such cadential, multi-wavelength data may lead
  to the first observations of compact SMBH binaries
  with sub-parsec separations.
\end{itemize}

We have assumed that a PTA source would appear
as luminous AGN, with the gaseous fuel perhaps
being supplied by the preceding merger of the binary's
host galaxies.
The degree to which galaxy mergers dictate AGN activity
remains an open question, and the link between
SMBH binarity and AGN activity is even less certain.
It is possible that many SMBH binaries that
are individually detected by PTAs will have no EM
counterpart at all.

Our results were calculated using a simple semi-analytic accretion disc
model, a central assumption of which is that the binary's tidal
torques are able to open a central cavity in the disc.
In the radiation-dominated regions of interest,
strong horizontal advective fluxes or vertical thickening of the disc
may act to close such a cavity and wipe out the features we predict.
The features would also not be present
if the disc and binary's orbits do not lie on the same plane,
as in the binary model of the variable BL Lac object OJ 287 \citep{LV96}.
Absorption and reprocessing by the binary's host galaxy
may also act to mask or mimic the intrinsic thermal AGN emission
we have modeled.

\section*{ACKNOWLEDGMENTS}
As we were completing this work, we became aware of a concurrent
independent study by \cite{Sesana+11}, addressing similar questions.
TT acknowledges fruitful discussions with
Alberto Sesana, Massimo Dotti, and Constanze R\"odig.
The authors thank Jules Halpern and Jeremy Goodman for insightful
conversations, and are grateful to the anonymous referee
for suggestions that improved the clarity of the manuscript.
This work was supported by NASA ATFP grants
NNXO8AH35G (to KM) and NNH10ZDA001N (to ZH)
and by the Pol\'anyi Program of the Hungarian National Office
for Research and Technology (NKTH; to ZH).
This research was supported in part by
the Perimeter Institute for Theoretical Physics.

\newpage
\begin{onecolumn}
\section*{Appendix: Green's-function solution for the thin-disc equation with moving inner boundaries}

Equation \ref{eq:sigt} states that given
an initial surface density profile $\Sigma_{\rm init}(R^{\prime})$,
the general solution $\Sigma(R,t)$
to the Keplerian thin-disc equation (equation \ref{eq:diff})
can be written
in the form $\Sigma(R)=\int G(R,R^{\prime})~\Sigma_{\rm init}(R^{\prime})~dR^{\prime}$.
The Green's function $G$ may be thought of as a
transform kernel that solves the thin-disc equation
for any choice of the time $t$, inner boundary $R_{\lambda}>0$,
the viscosity power-law $n\equiv \ln \nu/\ln R <2$
and the  ``leakage parameter'' $f_{\rm leak}$ (defined in equation \ref{eq:bc2})
that quantifies the mass flow rate across $R_{\lambda}$.
Below, we consider a case where where the boundary $R_{\lambda}$
moves inward as an explicit function of time.
For simplicity, we assume $n$ and $f_{\rm leak}$ to be constants.

Our astrophysical problem has two convenient properties that aid our mathematical analysis:
first, that the gas leaking into the cavity does so in nearly radial orbits
means it is effectively decoupled from the circumbinary disc
and the matter inside $R<R_{\lambda}$ can be ignored;
second, the disc is empty inside the initial value of $R_{\lambda}$
and so it is only necessary to integrate over $R^{\prime}\ge R_{\lambda,0}=R_{\lambda}(t=0)$.
Our present approach will not work, for example,
for a general boundary that moves outward with time.

Suppose that  $\mathcal{G}(R,R^{\prime}; t^{*})$ is a weighted sum
of the Green's function $G$ that consists of solutions evaluated
at different values of  $R_{\lambda}(t)$ over the duration $0\le t \le t^{*}$.
By the principle of superposition, $\mathcal{G}$ must
also be a solution of the thin-disc solution that satisfies
\beq
\Sigma (R,t^{*})=\int_{R_{\lambda,0}}^{\infty}
\mathcal{G}(R,R^{\prime};t^{*})
~\Sigma_{\rm init}(R^{\prime})~ dR^{\prime}.
\label{eq:sigev}
\eeq
At any $t$ and $R_{\lambda}$,
$\mathcal{G}$ must satisfy $\dd \mathcal{G}/\dd t = \dd G/\dd t$ 
and  $\dd \mathcal{G}/\dd R_{\lambda} = \dd G/\dd R_{\lambda}$.
Further, if $R_{\lambda}$ is an explicit function of time, it must be true
that the full time dependence of $\mathcal{G}$
is described by $d \mathcal{G}/dt = \dd \mathcal{G}/\dd t
+(\dd \mathcal{G}/\dd R_{\lambda}) (dR_{\lambda}/dt)$.
It then follows that
\begin{eqnarray}
\qquad
\Sigma (R,t+\Delta t) &=& \int_{R_{\lambda,0}}^{\infty}
\left[
\mathcal{G}(R,R^{\prime}; t, R_{\lambda})
+ \frac{d \mathcal{G}(R,R^{\prime}; t, R_{\lambda})}{d t}\Delta t
+\mathcal{O}(\Delta t^{2})+...
\right]\Sigma_{\rm init}(R^{\prime})~dR^{\prime}\nonumber\\
&=&
\int_{R_{\lambda,0}}^{\infty}
\left[
\mathcal{G}(R,R^{\prime}; t, R_{\lambda})
+ \frac{\dd G(R,R^{\prime}; t, R_{\lambda})}{\dd t}\Delta t
+ \frac{\dd G(R,R^{\prime}; t, R_{\lambda})}
{\dd R_{\lambda}} \frac{ d R_{\lambda}}{dt}\Delta t
+...
\right]\Sigma_{\rm init}(R^{\prime})~dR^{\prime}
.
\label{eq:sigtdt}
\end{eqnarray}

From equations \ref{eq:sigt} and \ref{eq:sigtdt},
we may write the derivative $d\Sigma/dt$
through its definition:
\beq
\frac{d}{dt}\Sigma (R,t)
=\int_{R_{\lambda,0}}^{\infty}
\left[\frac{\dd G(R,R^{\prime};t, R_{\lambda})}{\dd t}
+\frac{\dd G(R,R^{\prime};t, R_{\lambda})}{\dd R_{\lambda}}\frac{dR_{\lambda}}{dt}\right]
~\Sigma_{\rm init}(R^{\prime})~ dR^{\prime}.
\label{eq:dsigdt}
\eeq
Direct integration gives the expression
\beq
\Sigma (R,t^{*})=\Sigma_{\rm init}(R^{\prime})+\int_{0}^{t^{*}}\int_{R_{\lambda,0}}^{\infty}
\left[\frac{\dd G(R,R^{\prime};t, R_{\lambda})}{\dd t}
+\frac{\dd G(R,R^{\prime};t,R_{\lambda})}{\dd R_{\lambda}}\frac{dR_{\lambda}}{dt}\right]
\Sigma_{\rm init}(R^{\prime})~ dR^{\prime}~dt.
\eeq
Our ``master'' Green's function $\mathcal{G}$ is therefore
\beq
\mathcal{G}(R,R^{\prime}; t^{*})=\delta(R-R^{\prime})+\int_{0}^{t^{*}}
\left[\frac{\dd G(R,R^{\prime};t, R_{\lambda})}{\dd t}
+\frac{\dd G(R,R^{\prime};t, R_{\lambda})}{\dd R_{\lambda}}\frac{dR_{\lambda}}{dt}\right]
~dt.
\eeq
The Dirac $\delta$-function can be awkward to implement
in a numerical integration scheme. We rewrite it in terms
of the known Green's function with a fixed boundary by using the fact
that any Green's function evaluated at $t=0$ is the $\delta$-function,
i.e.,
\beq
G(R,R^{\prime}; t^{*}, R_{\lambda}^{*})=\delta (R-R^{\prime})
+\int_{0}^{t^{*}}\frac{\dd G(R,R^{\prime}; t, R_{\lambda}^{*})}{\dd t}~dt,
\eeq
where $R_{\lambda}^{*}=R_{\lambda}(t^{*})$.

We write our solution as
\beq
\mathcal{G}(R,R^{\prime}; t^{*})=G(R,R^{\prime}, t^{*}; R_{\lambda}^{*})
+\int_{0}^{t^{*}}\left[
  \frac{\dd G(R,R^{\prime}; t, R_{\lambda})}{\dd t}
  - \frac{\dd G(R,R^{\prime}; t, R_{\lambda}^{*})}{\dd t}
  +\frac{\dd G(R,R^{\prime}; t, R_{\lambda})}{\dd R_{\lambda}}\frac{dR_{\lambda}}{dt}
\right]
~dt.
\label{eq:Gmaster} 
\eeq

We see that if the boundary is stationary, the second term above
vanishes and $\mathcal{G}=G$.
Note the similarity of the mathematical form of our solution to DuHamel's
theorem \citep[e.g.,][]{CJ59} for time-dependent boundary conditions.
This is not surprising, given that both are based on the superposition principle;
equation \ref{eq:Gmaster} is simply a weighted sum of
instantaneous Green's functions and not, strictly speaking,
a new type of solution.
For our thin accretion disc problem,
the function $\mathcal{G}(R,R^{\prime},t)$ is explicitly known
and easily tabulated, given a specific combination
of the parameters $n$ and $f_{\rm leak}$
and the boundary evolution $R_{\lambda}(t)$.

In the context of this paper, the initial condition is the point when
the binary's orbital decay becomes GW-driven,
and $t^{*}$ is the time when the system is observed as an
individually resolved PTA source
and $R_{\lambda}(t)\ge R_{\lambda}^{*}$ describes the (GW-driven)
evolutionary history of the gap-opening radius prior to $t^{*}$.

\end{onecolumn}

\begin{twocolumn}
\bibliographystyle{mn2e}

\end{twocolumn}

\end{document}